\documentclass[showpacs,preprintnumbers,amsmath,amssymb,twocolumn,nofootinbib]{revtex4}
\usepackage{graphicx}% Include figure files
\usepackage{dcolumn}% Align table columns on decimal point
\usepackage{bm}% bold math
\usepackage{hyperref}
\usepackage{epsfig,slashed}
\newcommand{\nco}{\newcommand}
\nco{\beq}{\begin{equation}} \nco{\eeq}{\end{equation}}
\nco{\beqa}{\begin{eqnarray}} \nco{\eeqa}{\end{eqnarray}}
\nco{\lra}{\leftrightarrow}
\def\sfrac#1#2{{\textstyle{#1\over #2}}}

\nco{\sss}{\scriptscriptstyle} \nco{\dphi}{\varphi}
\nco{\lsim}{\mbox{\raisebox{-.6ex}{~$\stackrel{<}{\sim}$~}}}
\nco{\gsim}{\mbox{\raisebox{-.6ex}{~$\stackrel{>}{\sim}$~}}}

\def\sX{{\scriptscriptstyle X}}
\def\sS{{\scriptscriptstyle S}}
\def\sT{{\scriptscriptstyle T}}

\newcommand{\comment}[1]{}

\begin{document}

\title{Nonabelian dark matter: models and constraints}

\author{Fang Chen, James M.\ Cline,  Andrew R.\ Frey}

\affiliation{%
\centerline{Physics Department, McGill University,
3600 University Street, Montr\'eal, Qu\'ebec, Canada H3A 2T8}
e-mail: fangchen, jcline, frey\ @physics.mcgill.ca }

\date{July 2009}

\begin{abstract}  Numerous experimental anomalies hint at the
existence of a dark matter (DM) multiplet $\chi_i$  with small mass
splittings.  We survey the simplest such models which arise from DM
in the low representations of a new SU(2) gauge symmetry, whose gauge
bosons have a small mass $\mu\lsim$ 1 GeV.   We identify preferred
parameters $M_\chi \cong 1$ TeV, $\mu\sim 100$ MeV, $\alpha_g\sim
0.04$ and the $\chi\chi\to 4e$ annihilation channel, for explaining
PAMELA, Fermi, and INTEGRAL/SPI lepton excesses, while remaining
consistent with constraints from relic density, diffuse gamma rays
and the CMB.  This consistency is strengthened if DM annihilations
occur mainly in subhalos, while excitations (relevant to the
excited DM proposal to explain the 511 keV excess) occur in the galactic
center (GC), due to higher velocity dispersions in the GC, induced by baryons.   
 We derive new constraints and predictions which are
generic to these models.  Notably, decays of excited DM states
$\chi'\to\chi\gamma$ arise at one loop and could provide a new signal
for INTEGRAL/SPI; big bang nucleosynthesis (BBN) constraints on the
density of dark SU(2) gauge bosons imply a \emph{lower}
bound on the mixing parameter $\epsilon$ between the SU(2) gauge
bosons and photon.  These considerations rule out the possibility of the
gauge bosons that decay into $e^+e^-$ being long-lived.   We study in
detail models of doublet, triplet and quintuplet DM, showing that
both normal and inverted mass hierarchies can occur, with mass
splittings that can be parametrically smaller ({\it e.g.,} $O(100)$
keV) than the generic MeV scale of splittings.  A systematic
treatment of $Z_2$ symmetry which insures the stability of the
intermediate DM state is given for cases with inverted mass
hierarchy, of interest for boosting the 511 keV signal from the
excited dark matter mechanism.

\end{abstract}

\maketitle

\section{Introduction} 

In the last year, it was intriguingly suggested that a variety of
observed astrophysical anomalies might be tied together by a single
theoretical framework, in which transitions between states in a dark
matter (DM) multiplet, mediated by new GeV-scale gauge bosons, could 
lead to production of lepton pairs \cite{AH}.  These could explain
excess electron/positrons seen by the PAMELA \cite{pamela}, ATIC
\cite{atic}, PPB-BETS \cite{ppb-bets}, HEAT \cite{heat} and
INTEGRAL/SPI \cite{integral} experiments (the latter via the excited
DM proposal (XDM) \cite{xdm}).  In addition, it has been proposed
that such transitions could account for the DAMA/LIBRA annual
modulation \cite{dama} via the inelastic DM mechanism (iDM)
\cite{idm}.  Synchrotron radiation from the leptons could explain the
WMAP haze \cite{haze}.  More recently Fermi/LAT \cite{fermi} and
HESS  \cite{hess} have made higher precision measurements of the
$e^+e^-$ spectrum at TeV energies, confirming an excess above the
known background, although less  pronounced than the ATIC data.  The
DM explanation for this excess has by now been studied by  numerous
authors \cite{Delahaye,Cirelli:2008pk,Ibarra:2008jk,Yin:2008bs,Cholis,MPV,
Mardon:2009rc,Rothstein,Hamaguchi:2009jb,Malyshev,Berg,Balazs:2009wm,Meade,Liu}, 
and a plethora of models has been
proposed \cite{models}, including ones where the DM decays rather
than annihilates \cite{decays}. Pulsars provide a more conventional
astrophysical explanation\footnote{An even more conservative
interpretation is that no new source is needed to fit the data; see
for example ref.\ \cite{katz}, or concerning the 511 keV excess,
ref.\  \cite{lingenfelter}}  for many of these anomalies, but the
data do not yet clearly prefer them over the DM hypothesis
\cite{pulsars}.  However, constraints from secondary gamma rays
produced by the charged leptons (or from primary neutrinos)  are
rapidly closing up the allowed DM parameter space 
\cite{Hisano,Bergstrom:2008ag,Borriello:2009fa,Cirelli:2009vg,CyrRacine:2009yn,
Regis:2009md,Pinzke:2009cp,Hisano:2009fb,Spolyar:2009kx,Profumo,Slatyer,
Belikov:2009cx,Huetsi,Kuhlen,Cirelli,Cholis2,Kanzaki}. 
Anticipated new data from the Fermi
telescope is expected to tighten these constraints in the near
future.

The theoretical paradigm we focus on here assumes that the DM
transforms nontrivially under a nonabelian gauge symmetry which is
spontaneously broken below the 10 GeV scale.   Radiative corrections
from virtual gauge bosons induce mass splittings between the DM
states of order $\alpha_g\mu$, where $\alpha_g=g^2/4\pi$ is the fine
structure constant of the new gauge symmetry and $\mu \sim g v$ is a
characteristic gauge boson mass after spontaneous symmetry breaking. 
Multiple exchanges of the light gauge (or Higgs) bosons gives a 
Sommerfeld enhancement \cite{AH, Sommerfeld} which can explain the large
annihilation cross section needed in the galaxy, compared to the
smaller one in the early universe at the DM freeze-out temperature,
expected from the relic density. In our previous paper \cite{us}, we
presented an SU(2) model along these lines which was designed to more
easily give a large enough 511 keV signal as observed by INTEGRAL
while also accommodating the PAMELA/ATIC observations.  

Our goal in the present paper is to give a more comprehensive survey
of models based on SU(2) gauge symmetry, considering a few different
possibilities for  the means of coupling the DM to the standard
model, for the representation of the DM multiplet, and that of  the
scalars which break the gauge symmetry.  We also derive some new
constraints on the gauge and Higgs couplings which are particular to
this class of models. We start by discussing a number of general
issues which transcend the individual models.  

The paper is organized as follows.  Section \ref{km} details the
mechanism of kinetic mixing  of dark and standard model (SM) gauge
bosons, including its possible UV origin, and we derive a new
constraint on the gauge coupling from the induced DM transition
magnetic moment in the nonabelian case.  Section \ref{hm} discusses
the alternative of communication between the dark and SM sectors by
Higgs mixing. We derive new constraints on diagonal Yukawa couplings
of the dark Higgs to DM, from direct detection and from antiproton
production in the galaxy.  In section \ref{imh} we discuss the
concept of an inverted DM mass hierarchy for boosting the predicted
511 keV INTEGRAL signal, and the $Z_2$ symmetry and nonthermal DM
history needed to make this idea work.  Section \ref{fits} analyzes
which regions of parameter space best fit the experimental anomalies
(we do not insist on explaining DAMA, since the constraints on the
iDM mechanism have become so severe 
\cite{dama-constraints,dama-constraints2},\cite{BPR}) and
constraints from diffuse gamma  rays, relic density, big bang
nucleosynthesis, and laboratory constraints.  

In the remainder of the paper we discuss several
specific kinds of models, organized according to the SU(2)
representation of the DM.  Sections \ref{ddm}, \ref{tdm} and
\ref{qdm} respectively deal with DM in the doublet, triplet and
quintuplet representations.  In all of these models the gauge group
is simply SU(2).  For completeness and contrast, in section
\ref{su2u1} we consider one model with dark gauge group
SU(2)$\times$U(1) and triplet DM, which illustrates the differences
between the purely nonabelian models and ones where gauge kinetic
mixing occurs between U(1) field strengths.  We summarize our
findings in \ref{conc}.  Appendices \ref{tmm} and \ref{rmca}
respectively give details of the transition magnetic moment and
radiative mass computations, \ref{AppC} computes the annihilation
cross sections for freeze-out of DM in a general representation, and 
\ref{dmm} treats the diagonalization of the gauge boson and DM mass
matrices for the SU(2)$\times$U(1) model.

\section{\bf Kinetic mixing of gauge bosons}
\label{km}  A simple way of generating
couplings between one of the SU(2) gauge bosons and electrons is
through nonrenormalizable couplings of the form
\beq
	\sum_i{1\over\Lambda_i} Y_{\mu\nu} B_a^{\mu\nu}\Delta^a_{i}
\label{Bmixing3}
\eeq
or
\beq
	\sum_i{1\over\Lambda_i^2} 
	Y_{\mu\nu} B_a^{\mu\nu}h_{i}^\dagger \tau_a h_{i}
\label{Bmixing2}
\eeq
where $\Delta_a$ and $h$ are respectively triplet and doublet Higgs
fields which are assumed to get a VEV.   By having several triplet or
doublet fields (labeled by index $i$) which get VEV's in different 
directions, it is possible
to get mixing with several colors of the $B$ gauge boson. 
In (\ref{Bmixing3}), note that only a single linear combination of $B$ vectors
mixes with the SM.  With a generic Higgs potential, 
we can always choose the linear combination of triplets in
(\ref{Bmixing3}) to be $\Delta_1$.
However, depending on the Higgs potential, the vector that mixes with the
SM may be a linear combination of several $B$ mass eigenstates.

To understand the consequences of gauge boson mixing, it is useful to
start with a simple example in which a massive abelian boson $B$ mixes with the
photon.  The kinetic term is 
\beq
	-\frac14\left( F_{\mu\nu}F^{\mu\nu} + B_{\mu\nu}B^{\mu\nu} 
	-2\epsilon  B_{\mu\nu}F^{\mu\nu} \right) +\frac12\mu^2
B_\alpha B^\alpha
\eeq
Since the U(1) gauge symmetry of the photon is unbroken, it must
remain strictly massless.  This restricts the form of the
transformation which diagonalizes the kinetic term to 
\beq
	A_\mu = \tilde A_\mu + \epsilon \tilde B_\mu
\label{Amixing}
\eeq
Therefore all particles which couple to the photon acquire a coupling
of strength $\epsilon e$ to the massive $B$ gauge boson.

\begin{figure}[t] %\smallskip 
\centerline{\epsfxsize=0.45\textwidth\epsfbox{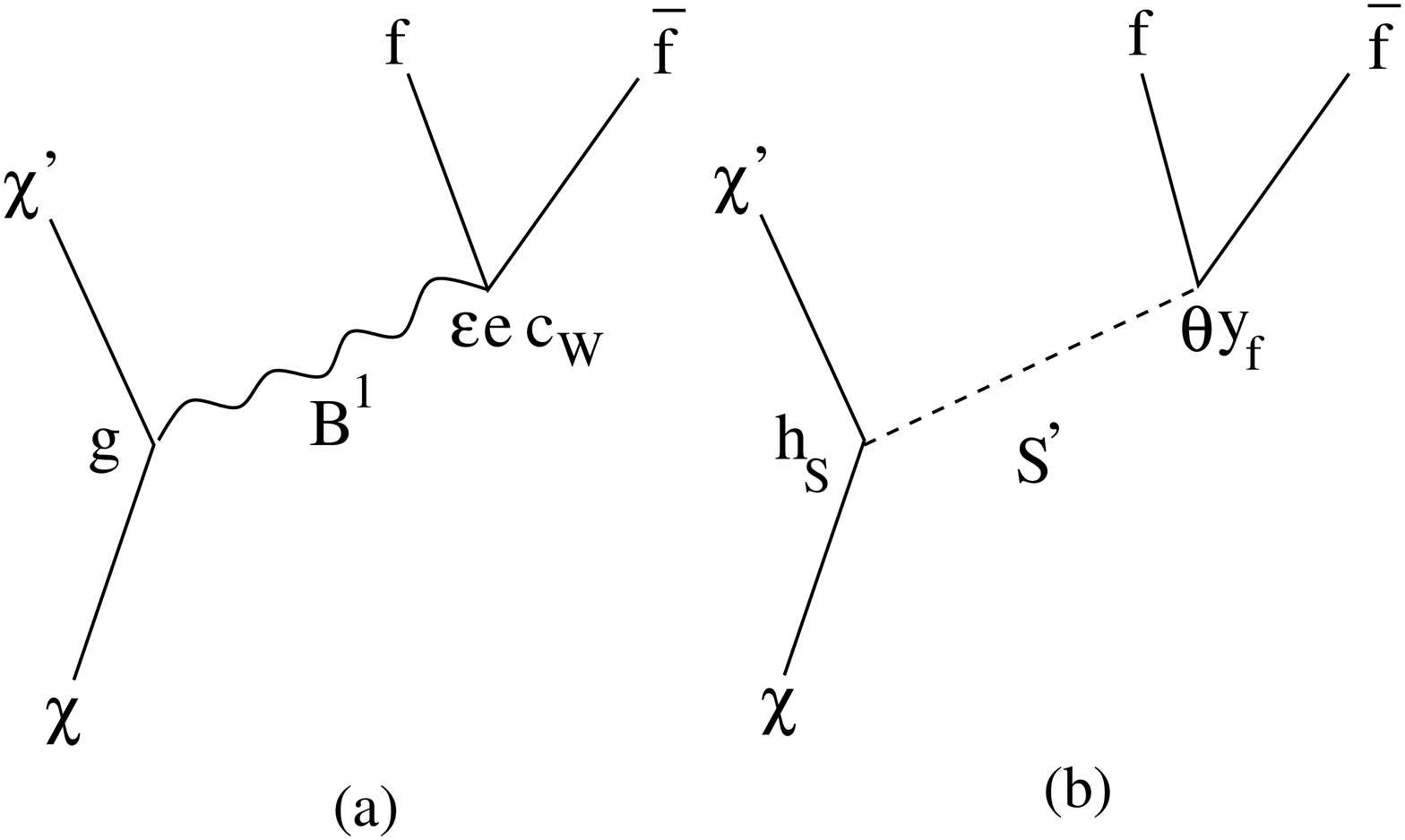}} 
\caption{Feynman diagrams for (a) $\chi\to\chi' f\bar f$ via gauge
boson mixing (left) or (b) Higgs mixing (right).} 
\label{diagrams}
\end{figure}

For the models we consider, the  mixing takes the form $\frac12\epsilon
Y_{\mu\nu} B_1^{\mu\nu}$, where for concreteness we take color 1
of the nonabelian gauge boson to mix with the standard model weak
hypercharge, it is straightforward to show that eq.\ (\ref{Amixing})
generalizes to the similar form
\beq
	A^\mu = \tilde A^\mu + \epsilon\cos\theta_W \tilde B_1^\mu
\label{Amixing2} 
\eeq
where $\theta_W$ is the Weinberg angle.  One must further transform
$B_1$ and the $Z$ gauge boson as
\beqa
	B_1^\mu &=& \left[\tilde B_1^\mu
	 - \epsilon\sin\theta_W 
	\left({m_Z^2\over m_Z^2 - \mu^2}\right)\tilde Z^\mu\right]
	\!\!\left(1+O(\epsilon^2)\right) \label{Bmix} \\
	Z^\mu &=& \left[\tilde Z^\mu + \epsilon\sin\theta_W \left({\mu^2 \over 
	m_Z^2-\mu^2}\right)\tilde B_1^\mu\right]
\!\!\left(1+O(\epsilon^2)\right)
\eeqa
where the tilded fields are those which diagonalize the kinetic
term.  Therefore the $B_1$ gauge boson acquires a coupling to the 
current of the $Z$ boson, in addition to that of the photon. 
Figure \ref{diagrams}(a) shows an example of a $\chi\to\chi' f\bar f$
transition mediated by the $B_1$.

The mass of the $Z$ gets shifted by a fractional amount
\beqa
	{\delta m_Z\over m_Z} = \epsilon^2\sin^2\theta_W {\mu^2\over 2
	m_Z^2} 
\eeqa
relative to its usual value.  For the small values of $\epsilon\sim
10^{-3}-10^{-4}$ and $\mu\lsim$ GeV which are of interest, this is a
negligible shift.

With gauge boson mixing, the annihilation $\chi\chi\to B_1 B_1$
results in subsequent decays of $B_1\to l^+ l^-$ with roughly
equal branching ratios for all leptons $l$ with mass below $\mu$.
Due to the nondiagonal couplings of $B_1$ to the $\chi$ states, 
assuming they are Majorana, 
there is no $s$-channel annihilation through a single virtual $B_1$.
Hence the annihilation into 4 leptons is guaranteed.  For Dirac DM,
such as in the doublet representation, this need not be the case,
as we will discuss in section \ref{ddm}.

\subsection{Microscopic origin of gauge kinetic mixing}
\label{epsilon}

\begin{figure}[b] %\smallskip 
\centerline{\epsfxsize=0.45\textwidth\epsfbox{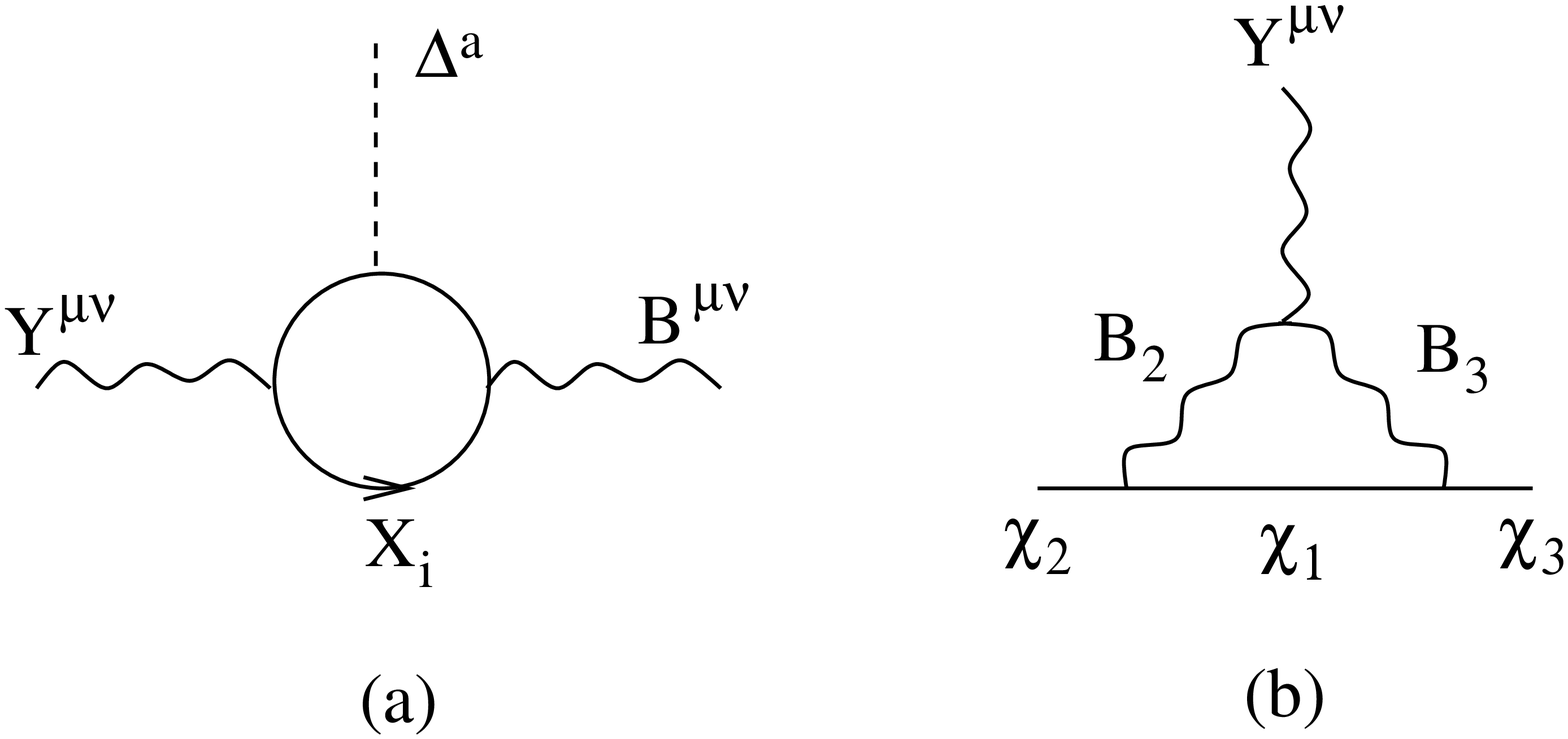}} 
\caption{Loop diagrams which generate (a) gauge kinetic mixing
(left) and (b) DM transition magnetic moment (right).} 
\label{loop}
\end{figure}

The dimension-5 operator (\ref{Bmixing3}) can be induced at one loop
by a heavy particle $X$ which carries both dark $SU(2)$ charge and weak
hypercharge $y_\sX$, if it also has a Yukawa coupling to the dark sector
Higgs triplet.  Suppose $X$ is a Dirac fermion which 
transforms as doublet of the SU(2), so
the Yukawa interaction is
\beq
	h_\sX \overline X_i (\tau_a)^i_j X^j \Delta^a
\label{heavyX}
\eeq
The diagram is shown in figure \ref{loop}(a). It generates the effective
interaction which can be estimated as
\beq
	 {h_\sX\, y_\sX\, g\over 16\pi^2 M_X} Y_{\mu\nu} B_a^{\mu\nu}\Delta_a
\eeq
so that the mixing parameter is given by $\epsilon\cong 
h_\sX y_\sX g \Delta / (16\pi^2 M_X)$, where $\Delta$ is the VEV of
the triplet Higgs.  For couplings of order unity and $\Delta\sim 10$
GeV, $M_X$ can be of order TeV to generate $\epsilon\sim 10^{-4}$.

Similarly, the dimension-6 operator (\ref{Bmixing2}) can arise from a heavy doublet
scalar field $S_i$ with a coupling 
%\beq
	$\lambda\, (S^\dagger\tau_a S)\, (h^\dagger \tau_a h)$
%\eeq 
to another dark higgs doublet $h$ (or perhaps the same one, $h\to S$).  If $S$ has weak hypercharge $y_\sS$,
the analogous diagram with $X$ replaced by $S$ gives rise to the
operator (\ref{Bmixing2}) with $\Lambda^2 \cong 16\pi^2 M_S^2 / 
(g \lambda y_\sS)$.
  
\subsection{Long-lived dark gauge bosons}
\label{llgb}

It is noteworthy that pure SU(2) models generically predict small
gauge mixing parameters $\epsilon$, suppressed by powers of a heavy
scale, whereas models with SU(2)$\times$U(1) gauge symmetry in the
dark sector allow for renormalizable mixing of SM and dark
hypercharge, in which case there is no reason to expect particularly
small values of $\epsilon$.  A phenomenological advantage of small
$\epsilon$ is that values on the order of $10^{-16}$ give the gauge
boson $B_1$ a lifetime of order $10^{12}$ s.  Such a long lifetime
lets $B$'s produced from DM annihilation  propagate away from the
galactic center before decaying.  This delocalizes gamma rays
produced by the leptonic decay products, allowing such models to
evade HESS constraints \cite{Rothstein, Meade}.  However we will show
in section \ref{rdbbn} that  gauge bosons with a lifetime greater than
$\sim\!1$ s are ruled out
by big bang nucleosynthesis for the models considered in this work.

\subsection{Direct decay of excited DM to photon}
\label{ddk}

Because there is no mixing of $B_1^\mu$ to $\tilde A^\mu$ in eq.\
(\ref{Bmix}), there is no tree level amplitude for the decay of
excited DM directly to a photon.  For example in the case of
triplet DM, one would have the decay $\chi_3\to\chi_2\gamma$ if
such a mixing existed.  Instead the dominant decay is 
$\chi_3\to\chi_2 l^+ l^-$ mediated by the $B_1$.  
However, in the class of models with kinetic
mixing between SM hypercharge and one of the dark SU(2) gauge bosons,
it is inevitable  for the single photon final state
to arise at the loop level, as we now show.  Naively, one could
draw the diagram where $l^+ l^-$ form a loop connecting $B_1$ to
the photon, but this just renormalizes the kinetic mixing term, so
it is not relevant.  There is another process which occurs due
to the nonabelian nature of the $B_1$, illustrated in fig.\ \ref{loop}(b).

The novel feature of the gauge mixing operator is that $B_1^{\mu\nu}$
contains the term $g(B_2^\mu B_3^\nu - B_2^\nu B_3^\mu)$.  There is
thus a trilinear vertex coupling these gauge bosons to the weak
hypercharge field strength, with strength $\epsilon g$.  One
consequence of this interaction is the generation of 
a transition magnetic moment for the DM.  An example is shown in
fig.\ \ref{loop}(b) for the case of DM in the triplet representation.
A magnetic moment interaction of the form
%\beq
	$\mu_{23}\bar \chi_2 \sigma_{\mu\nu} \chi_3 F^{\mu\nu}$
%\eeq
arises, where $\mu_{23}$ is expected to be of order 
$\epsilon g^3 /(16\pi^2 M_\chi)$.  A careful computation of the loop
diagram given in appendix \ref{tmm} gives
\beq
\label{MM}
	\mu_{23}\cong {\epsilon g^3 c_W\over 128\pi^2
M_\chi}\left(\ln{M_\chi\over\mu} -1\right)
\eeq
where $\mu$ is the scale of the nonabelian gauge boson masses and
$c_W=\cos\theta_W$.  It 
is straightforward to compute the rate $\Gamma_\gamma$ for
$\chi_2\to\chi_3\gamma$,
\beq
	\Gamma_\gamma = {\mu_{23}^2 \over 8\pi}(\delta M_{23})^3
\eeq
where $\delta M_{23} = M_{\chi_2}-M_{\chi_3}$ is the energy
available for the decay.  
On the other hand, the rate $\Gamma_{2e}$ for $\chi_2\to\chi_3 e^+
e^-$ is approximately $(\delta M_{23}-2 m_e)^2(\delta M_{23}
+2 m_e)/(256 \pi^3 M_\chi^2)$ times the spin-averaged 
squared matrix element, 
\beq
\langle |{\cal M}|^2\rangle 
\cong 32 g^2 e^2\epsilon^2 c_W^2 {M_\chi^2 \over \mu^4} \left(E_+E_- +\vec p_+\!\!\cdot\!\vec
p_- -m_e^2\right)
\eeq
(where $E_\mp$ and $\vec p_\mp$ are the energy and 3-momenta of the electron and positron,
respectively).  This varies approximately linearly over the allowed
phase space, so we estimate the integral as being 
\beq
	\Gamma_{2e} \cong 4\epsilon^2\alpha\alpha_g 
	(\delta M_{23}-2 m_e)^3(\delta M_{23} + 2 m_e)^2/ \mu^4
\eeq
The branching ratio for the single photon versus the two lepton
decay is thus
\beq
	{\rm BR}_\gamma = {c_W^2 \,\alpha_g^2 / \alpha\over 8192\pi^2}\,
	{\mu^4 (\delta M_{23})^3\over M_\chi^2 
	(\delta M_{23-})^3(\delta M_{23+})^2} \,
	\ln^2{M_\chi\over e\mu}
\eeq
where $\delta M_{23\pm}= \delta M_{23}\pm 2 m_e$ and 
$e = 2.71828\dots$.  Taking $\delta M_{23+}\cong 2\delta M_{23}
\cong 4 m_e$ but allowing for the possibility that $\delta M_{23-}\ll\delta
M_{23}$, we can write
\beqa
{\rm BR}_\gamma &\cong& 2.6\times 10^{-4}\, {\alpha_g^2\over\alpha}\,
	\left({\mu\over 200{\rm\ MeV}}\right)^4\left({1{\rm\ TeV}\over
	M_\chi}\right)^2
\nonumber\\ 	&\times& 
	\left({100 {\rm\ keV}\over \delta M_{23-}}\right)^3
\eeqa	
The reference values chosen here are compatible with constraints 
which we will discuss in later sections, and  small values of
$\delta M_{23-}$ enhance the size of ${\rm BR}_\gamma$.

Even though the branching ratio for $\chi_3\to\chi_2\gamma$ due to
the magnetic moment is small, the observable signal due to this
process, in the diffuse gamma ray background, is distinctive. If the
dark matter was at rest, it would produce a monoenergetic photon with
$E=\delta M\sim$ MeV.  Since the central galactic DM has a velocity
distribution with dispersion $v/c \sim 10^{-3}$, the spectrum of
the photon is Doppler broadened with a width of order 
$(v/c)\delta M  \sim 1$ keV for $\delta M_\chi\sim$ MeV.  This is
just below the $1.5$ keV resolution of SPI.  The
nonobservation of such a signal by INTEGRAL thus provides a new
constraint on  models with $S$-parameter type mixing of the
nonabelian gauge boson with weak hypercharge.

To determine the constraint, we can compare the new direct photon
signal with that of the 511 keV line already observed by INTEGRAL.
The latter is seen with a confidence level (c.l.) of 50$\sigma$ and 
a signal to background ratio ($S/B$) of a few percent.  One can
predict the c.l.\ of the new signal from that of the 511 keV line through
the relation
\beq
	({\rm c.l.})_{\rm new} = ({\rm c.l.})_{511}\,{\rm BR}_\gamma\, {(S/B)_{\rm new}\over
(S/B)_{\rm 511}}\left(\sigma_{511}\over 
\sigma_{\rm new}\right)^{1/2}
\eeq
where $\sigma_{511}\cong 5$ keV is the width of the 511 keV line and
$\sigma_{\rm new}= 1.5$ keV is the resolution of the detector
(which is approximately the same as the intrinsic line width).
To understand
the dependence on width, notice that for fixed flux, increasing the
width of a line reduces the signal proportionally ($1/\sigma$), but
for fixed signal-to-background,  it increases the counting statistics
by $\sqrt{\sigma}$ since  a wide line of a given intensity has more
flux than a narrow one.    These effects combine to give the
$1/\sqrt{\sigma}$ dependence.   The background for the 511 keV line
is dominated by the positronium $\to 3\gamma$ continuum and
annihilations of positrons in the INTEGRAL telescope, effects which
are both absent for the new signal.  On the other hand, there is a
broad instrumental line near 1.8 MeV which is the dominant background
for the narrow galactic $^{26}$Al line \cite{al26}, whose signal to  background
ratio is around 70/30.    Putting these numbers together, and assuming
that $({\rm c.l.})_{\rm new}<3$ to avoid a detection, we find the
limit
\beqa
	\alpha_g \lsim0.08\left({ 200{\rm\ MeV}\over\mu}\right)^{2}
	\left({M_\chi \over
	1{\rm\ TeV}}\right)\left({\delta
M_{23-}\over 100{\rm\ keV}}\right)^{3/2}
\label{alphacons}
\eeqa
(recall that $\delta M_{23-} = \delta M_{23}-2 m_e$).
It is interesting that such reasonable values of the dark gauge
coupling could lead to an additional signal potentially detectable by
INTEGRAL.  However, it would require a nonthermal DM history, since
we will show that smaller values of $\alpha_g$ are needed for the correct
relic density, eq.\ (\ref{fra}), or in the case of doublet dark
matter, the bound (\ref{alphacons}) does not apply because the
magnetic moment is suppressed by an additional factor of $\delta
M_{23}/M_\chi\sim 10^{-6}$, as we will show in section \ref{gkmd}.

\section{Mixing through the Higgs sector}
\label{hm}

\subsection{General features}

An alternative way in which the dark matter might couple to the standard model
is through renormalizable operators of the form
\beq
	\lambda_{\sss HS} |H|^2 |S|^2
\eeq
where $H$ is the standard model Higgs doublet and $S$ is a Higgs field
which is charged under the dark SU(2) gauge group.  If $S$ gets a 
VEV $v_S/\sqrt{2}$ 
and also has a Yukawa coupling to the DM, schematically of the
form $h_s S\chi\chi$, then transitions such as $\chi\to\chi f \bar f$
can be mediated by the Higgs bosons as shown in figure
\ref{diagrams}(b).
The Higgs sector has a mass matrix of the form
\beq
	\left(\begin{array}{cc} m^2_H & \lambda_{\sss HS} v_H v_S\\
	\lambda_{\sss HS} v_H v_S & m^2_S \end{array}\right)
\eeq
where $v_H$ is the VEV of the SM Higgs $h=\sqrt{2}H$.  If the mixing is small, then the
Lagrangian fields are related to the mass eigenstates by
\beq
	\left(\begin{array}{cc} H\\ S\end{array}\right)\cong
	\left(\begin{array}{cc} 1 & -\theta \\
	\theta  & \phantom{-}1 \end{array}\right) 
	\left(\begin{array}{cc} H'\\ S'\end{array}\right),\quad
	\theta = {\lambda_{\sss HS} v_H v_S\over m^2_H - m^2_S}
\eeq
Therefore $S'$ couples with strength $y_f\theta$ to any SM model
fermion $f$ whose Yukawa coupling to $H$ is $y_f$.   
In addition, the $H'$ couples to $\chi\chi$ with strength $-y_S\theta$. Thus the
diagram involving $H'$ exchange is of the same order in couplings
as that with the $S'$, but at low momentum transfer
it is suppressed by $m^2_S/m^2_H$.

\subsection{Constraints on diagonal couplings}
\label{cdc}

\subsubsection{No antiproton production}

\begin{figure}[t] %\smallskip 
\centerline{\epsfxsize=0.4\textwidth\epsfbox{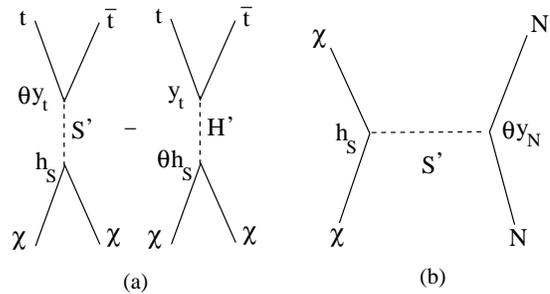}} 
\caption{(a) Left: DM annihilation into $t$-$\bar t$ by virtual
scalar that mixes with Higgs. (b) Right: DM scattering on nucleon by scalar exchange.
} \label{nucleon}
\end{figure}

An interesting qualitative difference between Higgs and gauge boson
mixing is that in the former case, the Yukawa couplings are generally
not off-diagonal.  For example, triplet dark matter coupling to a
quintuplet scalar as $\chi^a S_{ab}\chi^b$ has diagonal couplings;
similarly for doublet dark matter coupling to a triplet scalar via
$\chi_i \tau^a_{ij} \chi_j S_a$.  In either case, the annihilation
$\chi\chi\to S\to f^+ f^-$ shown in fig.\ \ref{nucleon}(a) occurs, resulting in quark or lepton pairs
favoring the most strongly coupled fermions---the top quark.  To
avoid production of hadrons, since no antiproton excess is observed
by PAMELA, one needs to have mixing with a scalar
that has dominantly off-diagonal couplings so that $\chi_1\chi_1$
annihilates  primarily to a pair of $S$ bosons by virtual
$\chi_2$ exchange.  The $S$ bosons decay nearly
on shell and hadron production can be suppressed if the $S$ is lighter
than $\sim\!1$ GeV.   Note that it is
impossible to keep the couplings strictly off-diagonal in the mass
basis, once the relevant component of $S$ gets a VEV, since this
contributes an off-diagonal mass term to the DM.  Therefore the Higgs
mixing scenario in its simplest form could be disfavored by the
lack of any antiproton excess in the PAMELA data.  

Moreover, diagonal
couplings are constrained by direct dark matter searches, by the
process shown in fig.\ \ref{nucleon}(b). 
Translating
the limit quoted in eq.\ (11) of ref.\ \cite{AH} to the present case
(and assuming $m_S =$ 200 MeV),  a diagonal Yukawa
coupling $h_s$ is bounded by
\beq
	\theta h_s y_N < 16\pi \times 10^{-8}\alpha_{\rm em}
\ \Rightarrow \ \theta h_s < 4\times 10^{-6}
\label{ddc1}
\eeq
Here $y_N \cong 10^{-3}$ is the Higgs-nucleon coupling \cite{Cheng}.
Assuming that the SM Higgs mass $m_H$ is much heavier than $m_S$, this
implies
\beq
	h_s \lambda_{\sss HS} < 2.7\times 10^{-4}
\left({m_S\over 200 {\rm\ MeV}}\right)^2 
\left({m_H\over 130 {\rm\ GeV}}\right)^2 
\left({1{\rm\ GeV}\over v_S}\right) 
\label{ddc2}
\eeq

To illustrate how severe (or mild) this constraint might be, consider the case of triplet DM $\chi_a$
coupled to a quintuplet (traceless symmetric tensor) scalar $S_{ab}$,
via $h_\sS \chi_a S_{ab} \chi_b$, and the cross-coupling $\lambda
H^2\, {\rm tr} S^2$ to the SM Higgs $H$.  Suppose for example that
$S_{12}$ gets a VEV $S\sim 1-10$ GeV to induce mixing with $H$, with
mixing angle $\theta\cong \lambda_{\sss HS} S v_H / m^2_H$.  In
addition to radiatively generated mass splittings of the DM (as we
will discuss below), there is a tree level contribution $h_\sS
S\chi_1\chi_2$ so that the mass eigenstates become linear
combinations, $\chi_\pm = \sqrt{1/2}(\chi_1\pm\chi_2)$.   The
fluctuations $\delta S$ of $S_{12}$ thus couple to the mass
eigenstates as $h_\sS  \delta S (\chi_+^2 - \chi_-^2)$.  Therefore
the ground state $\chi_-$ can  annihilate directly into a single
fermion pair through a single intermediate scalar.  The latter is
always far off shell, so this annihilation channel is dominated by
production of top quarks which hadronize and produce antiprotons,
contrary to the observations.   However notice that  the two diagrams in fig.\
\ref{nucleon}(a) interfere destructively.  We can estimate the effect
of these diagrams by integrating
out the intermediate scalar, and using the fact that $m_S\ll m_H$,
to get the effective dimension-6 operator
\beq
	\theta y_t h_\sS {m^2_H\over M^4}\, \chi\chi \, \bar t t
\eeq
On the other hand, the annihilation $\chi\chi\to S'S'$ by $\chi$
exchange can be estimated from the dimension-5 operator
\beq
	{h_\sS^2\over M_\chi} \chi\chi S'^2
\eeq
Assuming that the initial $\chi$'s are nonrelativistic, the ratio of
the corresponding cross sections is of order
\beq
	{ \sigma(\chi\chi\to\bar t t)\over \sigma(\chi\chi\to S'S')}
	\sim {\theta^2 y_t^2\over h_\sS^2}\, {m_H^4\over M_\chi^4}
\label{annrat}
\eeq
The top quarks decay to $b$ quarks before hadronization, and each
$b$ quark produces $\sim\!4.5$ antiprotons (using MicrOMEGAs
\cite{micro}), so the number
of antiprotons per positron is of the same order.  The observed flux
of antiprotons to electrons is approximately $10^{-3}$, and given that
no antiprotons in excess of standard expectations are observed, we
should demand that the ratio (\ref{annrat}) not exceed this limit.
For definiteness, if $M=1$ TeV  we
obtain the rather weak constraint
\beq
	\lambda_{\sss HS} \lsim 85\, h_\sS
\left({M_\chi\over 1{\rm\ TeV}}\right)
	\left({1{\rm\ GeV}\over v_S}\right)
\label{pbar}
\eeq 
Both (\ref{pbar}) and 
the direct detection constraint (\ref{ddc2})
can be satisfied using reasonable values of the couplings.

Furthermore, if there are additional contributions to the DM mass
splittings, it is possible to parametrically suppress the diagonal 
couplings.  For example, consider a second quintuplet Higgs $T_{ab}$ with
coupling $h_\sT\chi_a T_{ab}\chi_b$, and a VEV which splits the
$\chi$ masses diagonally, $h_\sT T(\chi_1\chi_1 - \chi_2\chi_2)$. In
this case, the $\chi$ mass eigenstates are not maximal mixtures of
the flavor states; rather $\chi_+ = \chi_1 + \delta\chi_2$, $\chi_- =
\chi_2 - \delta\chi_1$, with $\delta = h_\sS S/ h_\sT T$ (assuming
$\delta$ is small).  If the $|T|^2|H|^2$ coupling is negligible, then
the overall effect is to reduce the diagonal  couplings by the factor
$\delta$, while leaving the off-diagonal couplings unsuppressed.

The constraint due to the assumed lack of production of antiprotons
would be weakened even further if the recent claim of ref.\
\cite{Kane} is verified.  This work questions the assumption that the
observed antiproton background is actually understood in terms of
physics other than dark matter annihilation.  

\subsubsection{No two-lepton final states} 

In section \ref{dgrc} we will discuss the fact that recent
constraints on DM annihilation from the diffuse gamma ray background
are more severe for models in which $\chi\chi\to 2l$ than $4l$ final
states (where $l$ is a charged lepton) due to the harder spectrum in
two-body decays. This is not an issue when the intermediate particle
is an SU(2) gauge boson, since its couplings are automatically
off-diagonal and thus two bosons must be emitted in the annihilation,
but it might be an issue for intermediate Higgs bosons with diagonal
couplings.  However, the result (\ref{annrat}) can be directly
adapted to the case of decays to a lepton pair instead of a top pair
by substituting the lepton Yukawa coupling for that of the top.  Even
for the heaviest lepton, $\tau$, the result is suppressed by
$(y_\tau/y_t)^2 \cong 10^{-4}$.  These annihilations are thus much
more rare than those with the $\bar t t$ final states, and do not
provide a stronger constraint than the one derived above, even if we
only demand that the ratio be $\ll 1$ rather than $\lsim 10^{-3}$.  

\subsection{Long-lived dark Higgs boson}

In order to realize the long-lived intermediate state proposal of
ref.\ \cite{Rothstein}, it is interesting to know how small a mixing
angle is required to get the Higgs lifetime to be $10^{12}$ s. In
section \ref{fits} it will be aruged that
 Higgs masses in the range mass $m_{S}\lsim
100$ MeV are the most promising for fitting PAMELA/Fermi observations,
such that only the $e^+e^-$ final state is available.  Using
the decay rate $\Gamma \cong \theta^2 y_e^2 m_{S}/16\pi$, we find
that 
\beq
	\theta = 6\times 10^{-12} \left({100{\rm\ MeV}\over m_{S}}
	\right)^{1/2}
\label{thetaval}
\eeq
is the required value.  We will show in section \ref{crdgb-hm} that such
small values are strongly excluded by constraints on the
density of dark gauge bosons, which must decay before BBN.

\section{Inverted mass hierarchy and $Z_2$ symmetry}
\label{imh}

\begin{figure}[t] %\smallskip 
\centerline{\epsfxsize=0.45\textwidth\epsfbox{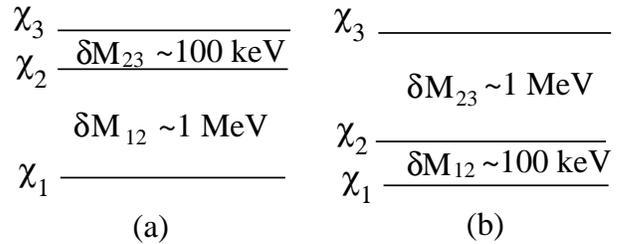}} 
\caption{(a) Left: inverted mass hierarchy of triplet DM;
(b) Right: normal hierarchy.} \label{mspect}
\end{figure}

In the following models, a recurring theme will be whether it is
possible to have a stable excited DM state which is slightly
lighter than the highest excited state (the one
that decays into leptons plus
ground state).  This ``inverted hierarchy'' is shown in figure
\ref{mspect}(a), in contrast to the ``normal hierarchy,'' fig.\
\ref{mspect}(b).  We proposed the inverted hierarchy in ref.\ 
\cite{us} as a means of boosting the galactic 511 keV signal from
excited dark matter, since the transition
$\chi_2\chi_2\to\chi_3\chi_3$ requires less energy than 
$\chi_1\chi_1\to\chi_3\chi_3$ and therefore benefits from a larger
proportion of the DM velocity distribution.

\subsection{Radiative mass corrections}
\begin{figure}[b] %\smallskip 
\centerline{\epsfxsize=0.45\textwidth\epsfbox{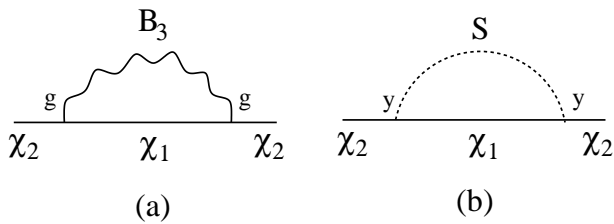}} 
\caption{Examples of radiative correction leading to mass splittings
within DM multiplet from exchange of virtual gauge bosons
(left) and Higgs bosons (right).} \label{self-energy}
\end{figure}

Let us first review the mechanism of radiative mass splitting
of a DM multiplet by virtual massive gauge bosons, through
diagrams like that shown in fig.\ \ref{self-energy}(a).  Although
the correction to the mass is logarithmically divergent, mass
differences between members of the multiplet are finite.  By choosing
a suitable counterterm, the finite part which contributes to the mass
splitting can be defined as
\beq
	\delta M_i \cong -\frac12\alpha_g \sum_j \mu_j T^j_{ia}T^j_{ai}
\label{rmc}
\eeq
where $\alpha_g = g^2/4\pi$ and the sum runs over all the gauge bosons,
with mass $\mu_j$, which contribute in the intermediate state. The
approximation (\ref{rmc}) is valid when $\mu_j \ll M_\chi$.
Details of the derivation are given in appendix \ref{rmca}.

If Higgs mixing rather than gauge boson kinetic mixing is the dominant
portal between the dark and SM sectors, it is likely that the dominant
souce of mass splittings is the tree level contributions from the
Higgs VEVs.  It is possible however that the 
analogous radiative corrections with the intermediate Higgses,
fig.\ \ref{self-energy}(b), have an
important effect.  In appendix \ref{rmca} it is shown that the
analogous formula to (\ref{rmc}) in this case is
\beq
	\delta M_i \cong +\frac14\alpha_{y} \sum_j m_j 
\label{rmh}
\eeq
where $y = \sqrt{4\pi\alpha_{y}}$ is the relevant Yukawa coupling
for the Higgs multiplet in the loop, and $m_j$ is the mass individual
components of that multiplet.

\subsection{$Z_2$ symmetry}

The idea of exciting  the intermediate state $\chi_2$ depends on it
being  significantly populated and  stable on cosmological time
scales.  One possibility is for it to be absolutely stable, which
should be guaranteed by some symmetry.  Another, which has been
explored in ref.\ \cite{FSWY}, is that the state is only metastable. 
In section \ref{llis} we will discuss that this scenario is strongly
constrained  by  direct detection considerations.   In this paper we
will highlight models that admit a  discrete $Z_2$ parity, which not
only ensures the stability of the intermediate state, but also
forbids transitions between it and the neighboring states, that could
be coupled to currents of SM particles.\footnote{Such transitions, if
they exist, can always mediate decays $\chi_2\to\chi_1 + 3\gamma$, as
in fig.\ \ref{loop3}(a).}   The absence of these transitions makes
the models safe from the direct detection constraints.  (For other
references discussing symmetries which stabilize DM, see
\cite{walker}.)

The simplest
example is triplet DM $\chi_i$ in which only one gauge boson color,
say $B_2$, mixes with the SM hypercharge.  In this case we can assign
conserved $Z_2$ charges  to the fields
\beq
	\chi_1,\ \chi_3, B_1, B_3
\eeq
and to no others.  Suppose that $\chi_1$
is the ground state and $\chi_3$
the heaviest state. Because of the $Z_2$ symmetry, $\chi_2$ can never
decay into $\chi_1$ plus SM particles.  It could in principle decay
into $\chi_1 B_3$, but this is kinematically blocked by the mass of the
$B_3$.
From the point of view of the symmetry, there is no light particle
that can appear in the final state to compensate the $Z_2$ charge of
the $\chi_1$.

Alternatively, we can state the condition that would make it
impossible to keep the intermediate state stable.  From the above
argument we see that a necessary requirement is to be able to
assign $Z_2$ charge to the ground state.  Therefore the highest
excited state of interest must also be charged.  If any gauge boson
which mediates transitions between the intermediate state and either
of the charged states mixes with SM hypercharge, then $Z_2$
charges cannot be consistently assigned.  

\subsection{$Z_2\times Z_2'$ symmetry for quintuplet DM}
\label{z2fq}

\begin{figure}[t] %\smallskip 
\centerline{\epsfxsize=0.45\textwidth\epsfbox{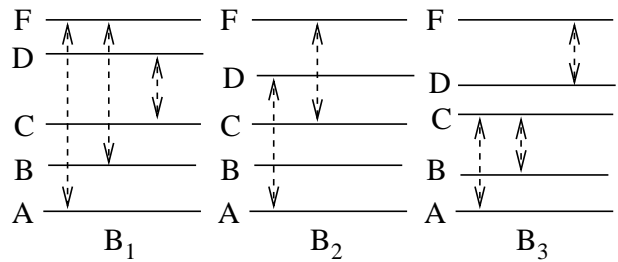}} 
\caption{Transitions between quintuplet states 
mediated by the three gauge bosons
$B_1$, $B_2$, $B_3$.} \label{5trans}
\end{figure}

The issue of having a stable intermediate state 
does not arise for DM in the doublet representation, but it
can be applied to higher representations, such as the symmetric
tensor (quintuplet).  We can label the canonically normalized 
states of $\chi_{ab}$ by
\beq
	\langle \chi\rangle = \left(\begin{array}{ccc}
	A-B/\sqrt{3}& C  & D\\
	C  & 2B/\sqrt{3} & F\\
	D  & F  & -A-B/\sqrt{3}\end{array}\right)
\label{5plet}
\eeq
The transitions mediated between these states by the three
$B_i$ gauge bosons are shown in figure \ref{5trans}.  Let us consider
how to assign $Z_2$ charges to the states in a systematic way.
First, suppose that one of the gauge bosons, say $B_a$, mixes with
SM hypercharge.  Then $B_a$ must not carry $Z_2$ charge, while the
other $B$'s do; call these $B_i$. This implies that some subset $X$ of 
$\chi$'s which appear only linearly and not bilinearly in the gauge 
interactions of the $B_i$'s should also be charged. The states in 
$X$ must also
have the property that they only appearly bilinearly and not linearly
in the interactions of $B_a$.  

Using this logic, we can make an exhaustive list of the possible 
$Z_2$ charge assignments for a given choice of the $B$ that mixes
with hypercharge, which we denote by $B_{a}\lra e^+ e^-$.  In the
process, we discover that actually the global symmetry is larger than
just $Z_2$; for a given subset $X$ of $\chi$ states, its complement
$Y$ could also have been chosen.  This means that we can assign
$Z_2$ charge to states in $X$, and a separate $Z_2'$ to states in $Y$.
Meanwhile, the two gauge bosons other than $B_a$ transform under both
$Z_2$ and $Z_2'$.  The result is 
\beqa
	B_{1}\lra e^+ e^-: &\ & X = \{C,D\},\ Y= 
	\{A,B,F\}\nonumber\\
	B_{2}\lra e^+ e^-: &\ & X = \{C,F\},\ Y = 
	\{A,B,D\}\nonumber\\
	B_{3}\lra e^+ e^-: &\ & X = \{D,F\},\ Y=
	\{A,B,C\}
\eeqa
It turns out that the $A$ and $B$ states always mix to form the
heaviest ($A'$) and lightest ($B'$) mass eigenstates.  We therefore
take the heaviest state  relevant for the INTEGRAL transition 
to be one in $Y$, and this dictates
that the intermediate state whose stability is to be guaranteed
is the lightest one in $X$.   The $Z_2\times Z_2'$ symmetry then 
insures that
the $Z_2$-charged intermediate state cannot decay into the 
$Z_2'$-charged lowest state,
since both symmetries would be violated.  
We will give explicit examples in section \ref{ihz2}.

In order for this to work, at least one of the $Z_2$'s must be left
unbroken by the VEVs of the Higgs fields.  If there is only one
triplet Higgs which gets a VEV to accomplish kinetic mixing, this
presents no difficulty since then the Higgs components can transform
in just the same way as the corresponding gauge fields, preserving
both $Z_2$'s.  Moreover  components of a quintuplet Higgs can be
given the same charges as the corresponding DM components, so one
$Z_2$ can be preserved as long as VEVs appear only in the $X$ or $Y$
subsets, but not both.  

VEVs of additional doublets break all of  the discrete symmetries,
but multiple triplet VEVs can be consistent with the symmetries if
they are orthogonal.  Consider two triplets with VEVs
$\vec\Delta$ and $\vec\Delta'$ in the 1 and 2 directions, respectively,
and suppose that $\vec\Delta$ is used to generate kinetic mixing
between $B_1$ and the SM. Then a single $Z_2$ is preserved, under
which the fields $B_2$, $B_3$, $\Delta_2$, $\Delta_3$ and
$\Delta'_1$ change sign, while $\Delta_1$, $\Delta'_2$ and $\Delta'_3$
do not.   Adding a third triplet $\Delta''$ with VEV in the 3
direction is also consistent with the $Z_2$, if $\Delta''_1$ 
transforms under it.

\subsection{Nonthermal history}

Even though $Z_2$ symmetry guarantees the stability of the
intermediate state $\chi_2$, it cannot prevent depletion of its
density in the early universe,  through exactly the same process
needed for the INTEGRAL signal, namely $\chi_2\chi_2\to\chi_3\chi_3$
followed by $\chi_3\to e^+ e^- \chi_1$ decay.   Even more simply, 
the depletion could occur directly by $\chi_2\chi_2\to\chi_1\chi_1$.
 In ref.\ \cite{us}, we
noted that this depletion could be prevented if the $\chi$'s were
produced out of thermal equilibrium rather than through the standard
freeze-out.  If the $\chi$'s are decay products of a supermassive
scalar $S$, their initially high energies suppress the annihilation cross
section sufficiently long to keep the $\chi_2\chi_2\to\chi_3\chi_3$
excitation or the $\chi_2\chi_2\to\chi_1\chi_1$ relaxation
out of thermal equilibrium in the early
universe.

In more detail, suppose that the gauge coupling $\alpha_g$ is too
large to yield the right relic density from freeze-out.  It was
envisioned that $S$ could decay at a low temperature $\sim\!5$ MeV,
resulting in mildly relativistic DM with momenta $p\sim 10^5 T$.  The
Sommerfeld enhancement is initially absent for DM with such large
velocity, and in fact the rate of annihilations remains always less
than the Hubble rate before cosmological structure begins to form,
because $n\langle\sigma v\rangle$ and $H$ both scale like $T^2$. 
Only when DM begins to concentrate in halos does the rate of
annihilations become significant.  

\section{Fitting PAMELA/Fermi/HESS versus INTEGRAL/SPI, cosmology and
laboratory bounds}
\label{fits}

\subsection{Fits to PAMELA/Fermi/HESS}
\label{pamela}

\begin{figure}[t] %\smallskip 
\centerline{\epsfxsize=0.4\textwidth\epsfbox{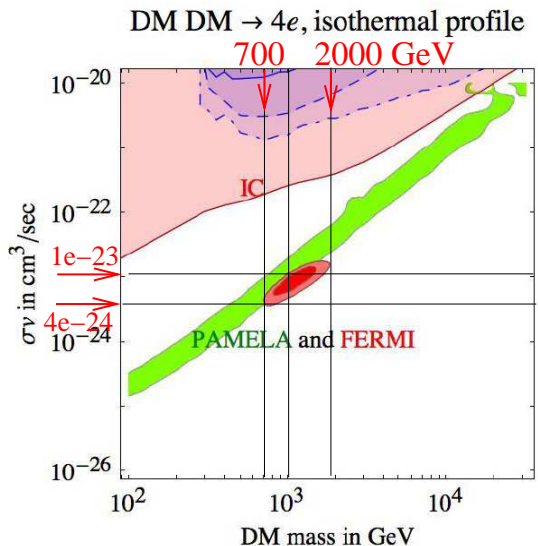}} 
%\centerline{\epsfxsize=0.4\textwidth\epsfbox{test.eps}} 
\caption{Best fit of ref.\ \cite{Meade} to the $\chi\chi\to 4e$
annihilation channel, in plane of $\langle\sigma v_{\rm rel}\rangle$
and $M_\chi$. Shaded regions in upper part are excluded by
diffuse gamma ray constraints.} \label{meade-fig}
\end{figure}

Ref. \cite{Meade} has identified regions in the parameter space of
$M_\chi$ and $\sigma_{\rm ann} v_{\rm rel}$ for the process
$\chi\chi\to BB$ (followed by $B\to e^+e^-$) which are compatible
with the PAMELA/Fermi/HESS $e^+e^-$ observations, as well as the HESS
constraints on inverse Compton gamma rays produced by the electrons
and positrons coming from DM annihilation.\footnote{A recent analysis
of preliminary Fermi observations of gamma rays from the inner galaxy
is also consistent with this annihilation channel \cite{Cholis2}.}
As we will discuss in
further detail below, additional constraints from extragalactic
diffuse gamma ray production favor the models in which the $B$'s from
$\chi\chi$ annihilation decay only to $e^+e^-$ and no heavier
leptons.   This implies that the mass of the $B$'s, $\mu$, must be
less than twice the mass of the muon, $\mu\lsim 200$ MeV.   The
allowed region for this scenario is reproduced in  fig.\
\ref{meade-fig}.  It should be emphasized that the two-body decay
$\chi\chi\to e^+e^-$ mediated by a single $B$ exchange is excluded
because its electron spectrum ends too abruptly due to its near
monoenergeticity \cite{Meade}; this channel also provides a very poor
fit to the PAMELA data \cite{MPV}. Moreover in the class of models considered
here, it would be impossible to forbid the channels $\chi\chi\to f\bar
f$ where $f$ is any SM fermion, if $\chi\chi\to e^+e^-$ is
unsuppressed.

The best fit is in the vicinity of
$\langle\sigma_{\rm ann} v_{\rm rel}\rangle\cong 10^{-23}$ cm$^3$/s
and $M_\chi\cong 1$ TeV.  This cross section exceeds that needed for
the correct thermal relic density ($10^{-36}\cdot c$ cm$^2$
\cite{PDG}) by a factor of $B=330$, which is thus the required boost
factor, assuming a thermal origin for the  DM.  Even if a nonthermal
origin is assumed, the thermal component  should be suppressed by
having an even larger cross section, and thus 330 should be regarded
as an upper bound on the required boost factor.

The example shown assumes an isothermal radial density profile, which
eases the constraints from HESS on the inverse Compton photons by
lowering the DM density near the galactic ridge.  For preferred
profiles such as Einasto, the fit to PAMELA/Fermi is nearly ruled
out. The isothermal profile is considered to be unrealistically flat
near the center compared to the results of the best N-body
simulations, but it was noted in \cite{Meade} that long-lived
intermediate bosons (the $B$ gauge bosons in our case) could justify
such an effective profile, due to the $B$'s traveling away from the
galactic center before decaying \cite{Rothstein}.  We will show that
in section \ref{bfd} below that big bang nucleosynthesis constraints
rule out such a long-lived $B$ in the present class of models, hence
the mechanism of long-lived intermediate states cannot work here.

Another way of decentralizing the region of DM annihilation has been
proposed in \cite{AH}, however, which could have a quantitatively
similar effect to the softer halo profile; namely DM subhalos which
populate the halo could dominate as annihilation sites, due to their
lower velocity dispersion and hence larger Sommerfeld enhancement.
The small-velocity subhalo scenario has recently been studied in
detail in ref.\ \cite{Kuhlen} (see also 
\cite{Robertson,Bovy:2009zs,vialactea}), with reference to models favored by
the pre-Fermi analysis of \cite{MPV}, in particular with $M_\chi=1$
TeV, $\mu=200$ MeV and $\alpha_g\cong 0.04$.   This happens to be
close to the preferred values mentioned above; we will show in the
next section that this value of $\alpha_g$ is just slightly larger
than the one needed to get the right relic density for triplet DM. 

Still, to avoid the stronger inverse-Compton constraints on the
preferred Einasto profile, it may be necessary to reduce the
annihilation rate near $r=0$, in addition to providing alternative
subhalo regions for the annihilation.  Recent work on halo formation
including the effects of baryons indicates that the velocity profile
steepens considerably (diverging like $r^{-1/4}$) for $r\lsim 20$ kpc
instead of leveling off to smaller values \cite{RD} as in pure DM
simulations.  (This reference also finds that the DM density profiles
are softened near the center, a result not  corroborated by other
simulations which include baryons \cite{Abadi}, but the latter work
does qualitatively confirm the steepening of the velocity profile
\cite{julio}.)  Moreover the overall magnitude of the velocity is
somewhat increased  for $r\lsim 100$ kpc.  Because the Sommerfeld
enhancement of the annihilation cross section scales like $1/v$, this
should have a similar effect to erasing the cusp of the density
profile, making it more similar to the isothermal profile.

\subsection{Relic density}
\label{rds}

We have computed the early-universe annihilation cross section of 
DM in any SU(2) representation into dark gauge bosons.
For the three representations we focus on in this paper, the result
is 
\beq
	\langle \sigma_{\rm ann}v_{\rm rel}\rangle \cong 
{\pi\alpha_g^2\over M_\chi^2}\times \left\{\begin{array}{cl}
	0.14, & \hbox{doublet}\\
	0.88, & \hbox{triplet}\\
	5.18, & \hbox{quintuplet}\end{array}\right\}
\label{sigann}
\eeq
Details are given in appendix \ref{AppC}.  
Using the standard value
$\langle\sigma v_{\rm rel} \rangle =10^{-36}$ cm$^2\cdot c$ needed
for thermal relic abundance \cite{PDG}, comparison with the cross
section (\ref{sigann}) indicates that the values of $\alpha_g$
required are
\beqa
\alpha_g &=& 
\left\{\begin{array}{cl}
	0.077, & \hbox{doublet}\\
	0.031, & \hbox{triplet}\\
	0.013, & \hbox{quintuplet}\end{array}\right\}
\times\left({M_\chi\over 1{\rm\ TeV}}\right)\\
&&\hbox{ (relic density value)}\nonumber
\label{fra}
\eeqa

As mentioned above, there are motivations to question the assumption that DM has
a thermal origin, such as our inverted mass hierarchy proposal 
\cite{us} (see also \cite{FSWY} and \cite{brand}).  It is important
to notice that to justify a nonthermal origin, the thermal
contribution must be smaller than usual so that it is subdominant to
the nonthermal contribution; thus the annihilation
cross section would be larger.  The values (\ref{fra}) should then
be regarded as lower bounds.  

These bounds can be evaded if the DM has stronger Yukawa couplings to
dark Higgs fields; for example triplet DM can have the coupling
$h\chi_i\Sigma_{ij}\chi_j$ to a quintuplet Higgs $\Sigma$.   If $h\gg
g$ then the freeze-out density is determined by $h$ and the gauge
coupling can be smaller than in (\ref{fra}).  In such a case, it should be kept in
mind that the annihilation in the galaxy will probably also be
dominated by Higgs boson exchange; notice that the mass scale of the
Higgs bosons cannot naturally exceed that of the gauge bosons by a
large factor, since the
scale of spontaneous breaking of the dark SU(2) gauge symmetry is 
dictated by the mass scales in the Higgs sector.  Thus late-time
annihilations would likely be dominated by Sommerfeld-enhanced Higgs
exchange diagrams.  The  expected boost factor would thus still be
$\sim\!300$ even in cases where $\alpha_g$ is much smaller than
indicated in (\ref{fra}).

\subsection{Mass splittings and the XDM (iDM) mechanism}
\label{xdmidm}

In contrast to the above values of $\alpha_g$, the paradigm of 
ref.\ \cite{AH} would at first seem to suggest smaller values
$\alpha_g\sim 10^{-3}$, because the radiative mass splittings 
of the DM multiplets go like $\alpha_g\mu$ (where $\mu$ is the scale
of the gauge bosons masses) and it was presumed that $\mu\sim 1$ GeV
as the largest value compatible with no production of antiprotons by
the decays of the gauge bosons after $\chi\chi\to BB$ annihilation in
the galaxy.  Since the XDM hypothesis requires $\chi$ mass splittings
of order MeV, $\alpha = $MeV/GeV$\sim 10^{-3}$ would be indicated.

However we have argued above that lighter gauge boson masses $\mu
\sim 100$ MeV are in better agreement with gamma ray constraints. The
generic estimate $\delta M \cong \frac12 \alpha_g\mu$ gives $\sim\!2$
MeV for such masses and the preferred coupling $\alpha_g\cong 0.04$
from section \ref{pamela}. 
This is in just the right range for having excited DM states which
can decay to $e^+e^-$ and the ground state.

For other applications, like the iDM mechanism for DAMA, or our
inverted mass hierarchy variant of XDM \cite{us}, it is desirable to
have splittings which are perhaps smaller than the MeV scale. In
section \ref{ortho} we will show that with sufficiently complicated
Higgs sectors (three triplets in this example) it is possible to
reduce the mass splittings below the generic level of $\alpha\mu$. 
It is also possible to design the gauge symmetry breaking (by
appropriate choices of VEV's or the DM representation) so that no
$\chi$ mass splittings are induced by gauge boson radiative
corrections; for doublet dark matter this is true regardless of the
Higgs respresentations.  In that case, the splittings must come from
Yukawa couplings and then it is possible to decouple the scale of the
gauge boson masses from that of the splitting.

It should be emphasized that getting the excited dark matter (XDM)
mechanism to produce a large enough signal to explain the INTEGRAL/SPI
observations is not as easy as just having the right DM mass 
splitting; one must generically saturate partial wave unitarity
bounds for the excitation cross section to get a large enough rate
\cite{PR}.  We leave the details of reanalyzing this problem to
work in progress \cite{new}.  The same can be said (even more so)
of the iDM mechanism for DAMA.  The region of parameter space
consistent with the DAMA annual modulation as well as other direct
detection experiments is essentially excluded \cite{dama-constraints2},
\cite{BPR}.
We give less emphasis to trying to implement the iDM mechanism.

\subsection{Overcoming diffuse gamma ray and CMB constraints}
\label{dgrc}

We have already seen that  constraints from gamma rays originating
as brehmsstrahlung or inverse Compton scattering of the emitted
leptons can often rule out models which would have provided good fits
to the PAMELA and Fermi observations \cite{Hisano}-\cite{Kuhlen}.  Not only
annihilations within our own galaxy provide such constraints, but
the accumulated effect from early redshifts and other halos on the
CMB and diffuse gamma ray background can be severe.  For example,
ref.\ \cite{Slatyer} obtains the 95\% c.l.\ CMB bound 
\beq
	\langle\sigma v_{\rm rel}\rangle < 4-8\times 10^{-24}
	{\rm cm}^3/{\rm s}
\eeq
for the model with $\chi\chi\to 4e$ and $M_\chi = 1-2$ TeV 
(where their efficiency factor
$f$ for transfering energy to the intergalactic medium is
approximately $0.9$).  This is barely compatible with the fit to 
PAMELA/Fermi/HESS for the same model in ref.\ \cite{Meade}, reproduced
in fig.\ \ref{meade-fig}.  

Many papers which place gamma ray constraints on annihilating DM
assume that only two leptons are produced, instead of the four which
are predicted by the class of models we are considering.  Given that
the preferred models are near the borderline of being excluded,
subject to large astrophysical uncertainties, the distinction between
the relatively hard, monoenergetic input spectrum for two-lepton annihilations
versus the softer four-body final states is important. In particular,
ref.\ \cite{MPV} (see section 4.1.3) has quantitatively shown this to be the case.   

Furthermore, in excluding a given model, one should keep in mind the
correlation between the best fit model parameters (the DM mass,
annihilation cross section, and gauge boson decay branching ratios)
with the assumed DM galactic density profile, since varying the
latter can cause significant changes in the former.  For example some
papers  refer to best-fit models as determined by ref.\ \cite{Meade},
but use different DM profiles to compute the constraints than those
used to fit the PAMELA/Fermi data, making it unclear which models are
really ruled out.

\subsection{Relic dark gauge (or Higgs) bosons and big bang nucleosythesis}
\label{rdbbn}

In this section we consider cosmological constraints on the lightest
stable or metastable particle in the dark sector.  Since we have
identified the mass scale $\mu\cong 100$  MeV for the portal boson
as being favored by fits to the PAMELA/Fermi/HESS data, we will take
this to be the lightest particle, be it the gauge boson in the case
of gauge kinetic mixing, or a Higgs boson in the case of Higgs mixing.
By this assumption we avoid the introduction of any scales which are
even lower than 100 MeV.  

Some of the bounds we derive implicitly assume that the dark gauge
bosons were in equilibrium with the rest of the plasma at a high
temperature, so that their abundance is known around the time when
they are becoming nonrelativistic.  Even if the mixing parameter
$\epsilon$ is too small for interactions with electrons to achieve
thermal equilibrium with the dark sector, one should remember that
kinetic mixing arises from some higher scale physics, such as a heavy
$X$ particle which transforms under both the dark and the SM gauge
symmetries; recall eq.\  (\ref{heavyX}).  Even for small values of
$\epsilon$, such an origin for the kinetic mixing can insure
equilibrium between the dark and SM sectors at the TeV scale.

\subsubsection{Long-lived gauge bosons}
\label{bfd}

In previous sections, it was noted that dark gauge bosons with long
$\sim\!10^{12}$ s lifetimes could have provided an escape from gamma ray
constraints on annihilating DM through the mechanism of ref.\ 
\cite{Rothstein}, but we now argue these would also dominate the
energy density of the universe at the time of BBN, assuming the DM
was produced thermally.  Let us consider the least dangerous case of
$\mu= 10$ MeV gauge bosons. Further, suppose that the SM becomes
supersymmetric above the weak scale, so that the number of degrees of
freedom is doubled; if instead there is a desert of no new states,
this will only make the BBN constraint stronger.  When the DM
particles freeze out between $T=M_\chi$ and $M_\chi/20$, they
transfer their entropy to the dark gauge bosons.  This increases
the energy density of the latter by at most a factor of two, since
there are more gauge degrees of freedom than DM ones.  In the
meantime, between temperatures of 1 TeV and $\mu=10$ MeV, the SM
degrees of freedom are differentially heated relative to the dark
gauge bosons by a factor of approximately 
$(214/11/2)^{1/3} \cong (9.7)^{1/3}$, due
to the change in the number of degrees of freedom from 214 to 11, and
the fact that the gauge bosons had been heated by a factor of $\sim\!2$
by  the DM annihilations.  (The precise value depends on the dimension
$d_R$ of the DM representation, but for the small-$d_R$ models we 
consider, this has no effect on the ensuing bound.)
Thus at $T=10$ MeV, the energy density in dark
gauge bosons is suppressed by a factor of $(9.7)^{4/3}\cong 21$ per
degree of freedom. By $T=1$ MeV this suppression has gone down to
a factor of 2.1 due to the gauge bosons being nonrelativistic. 
However there are 3 colors and 3 polarizations, so this counts as
approximately 4.5 extra species, and is ruled out.   We conclude that
the gauge boson lifetime should be less than 1 s (the time
corresponding to $T=1$ MeV), requiring that
\beq
	\epsilon > 4\times 10^{-11} \left({100{\rm\ MeV}\over\mu}
	\right)^{1/2}
\label{bbnc}
\eeq
We used the decay rate $\Gamma \cong \frac13\alpha\epsilon^2\mu$ for
$B\to e^+e^-$.

Even if the thermal relic DM density is highly depleted by having a
large annihilation cross section, the above arguments hold, since
most of the energy of the original thermal DM population is deposited
in the gauge bosons, regardless of how much DM is left.  The only
obvious way to avoid the above constraint on $\epsilon$ is to somehow
dilute the original DM even more relative to the SM, {\it e.g.,} by
having even more extra degrees of freedom present at a TeV than in
the minimal supersymmetric standard model. We note that the $B$
bosons will not equilibrate with the SM for values of $\epsilon$
lower than (\ref{bbnc}), so equilibration cannot serve to dilute the dark
gauge bosons.

\subsubsection{Stable gauge bosons}
\label{crdgb-gkm}

Typically only one color of the dark gauge bosons mixes with the SM,
say $B_1$, while transitions between $B_2$ and $B_3$ can be mediated
by the nonabelian mixing interaction $g\epsilon\cos\theta_W F_{\mu\nu}B_2^\mu
B_3^\nu$ which we referred to previously in section \ref{ddk},
leaving the lighter of these two states stable against decay.  We must
verify that its relic density is not too large.  

\begin{figure}[t] %\smallskip 
\centerline{\epsfxsize=0.3\textwidth\epsfbox{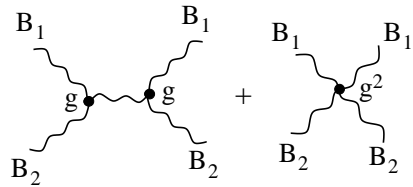}} 
\caption{Annihilations which deplete density of stable $B_2$ bosons.}
\label{bfreeze}
\end{figure}

For definiteness, suppose the stable gauge boson is $B_2$.  The most
efficient process for depleting $B_2$ is the scattering $B_2 B_2\to
B_1 B_1$, shown in fig.\  \ref{bfreeze}, followed by the decays $B_1
\to e^+ e^-$.  We will show that this is true even if $B_1$ is
heavier than $B_2$. 

\begin{figure}[t] %\smallskip 
\centerline{\epsfxsize=0.4\textwidth\epsfbox{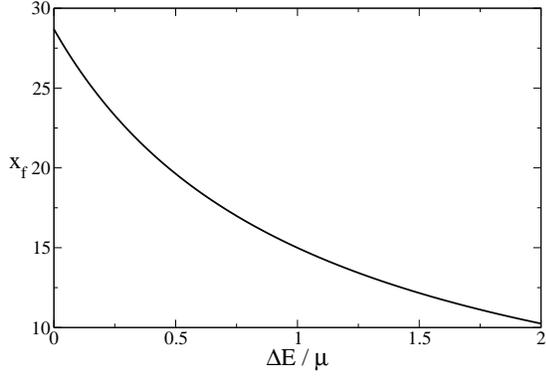}} 
\caption{Freeze-out value $x_f$ versus $\Delta E/\mu$ (solution of
eq.\ (\ref{xfeq})) for the process $B_2 B_2\to B_1 B_1$, where
$\Delta E = $ max$(2(\mu_1-\mu_2),0)$.}
\label{xffig}
\end{figure}

The cross section for $B_2 B_2\to B_1 B_1$ can be estimated as 
\beq
\langle\sigma v\rangle \sim e^{-\Delta E/T}\,{\alpha_g^2\over \mu^2}
\left({\delta \mu\over \mu}\right)^{1/2}
\label{sigvel}
\eeq
where $\Delta E$ is the energy barrier: $\Delta E=0$ if $\mu_2>\mu_1$,
and $\Delta E=2\delta\mu = 2(\mu_2-\mu_1)$ if  
$\mu_1>\mu_2$.  The factor of $(\delta\mu/\mu)^{1/2}$ arises from the
velocity of the final state particles, which is $1$ in the more
familiar case of annihilation to light final states.  The freeze-out
temperature for this reaction is determined as usual by setting 
$n_{B_2}\langle\sigma v\rangle$ equal to the Hubble rate, using the
equilibrium density of a massive particle for $n_{B_2}$; one finds
that
\beqa
	x_f ={\mu\over T_f} &=& {\ln\left(0.04 {\alpha_g^2\over\sqrt{g_*}}{m_P\over\mu}
	\right) - 2\ln x_f\over 1 + \Delta E/\mu}\nonumber\\
&\cong& {35.4 - 2\ln{x_f}\over 1 + \Delta E/\mu}
\label{xfeq}
\eeqa
for $\mu= 100$ MeV and $\alpha_g=0.04$.
This implicit equation quickly converges to a solution by iteration. 
Values of $x_f$ as a function of $\Delta E/\mu$ are shown in figure
\ref{xffig}.

As long as the interactions of fig.\ \ref{bfreeze} are in
equilibrium, the abundance $Y_{B_2}$ tracks that of $Y_{B_1}$, whose
principal connection with the SM is through the decays and inverse
decays $B_1\lra e^+e^-$.  The decay rate is suppressed by
$\epsilon^2$, and for the small values of $\epsilon$ we obtain  in
the ensuing bound, it is consistent to neglect scattering processes
$B_1 B_1\lra e^+e^-$ whose rate goes like $\epsilon^4$.  In appendix
\ref{3b2b} we show that the processes $B B B\to B B$ are able to keep
the gauge bosons in kinetic equilibrium with themselves down to a
temperature given by $x_k = \mu/T_k = 17.5$, so $B_1$ would maintain 
the equilibrium
abundance of a nonrelativistic particle $Y_{\rm eq}$ until this
temperature.    At lower temperatures, it disappears due to its
decays:
\beq
	Y_{B_1} =  Y_{\rm eq}(x_k) e^{-\Gamma t} = Y_{\rm eq}(x_k) e^{-(\Gamma/H(\mu))x^2} 
\label{yb1eq}
\eeq 
where $\Gamma =
\frac13\alpha\epsilon^2\mu$ is the decay rate, $x=\mu/T$, and
$H(\mu)$ is the Hubble rate at $T=\mu$.  Since $Y_{\rm eq}(x)\sim
x^{3/2}e^{-x}$, we find that $Y_{\rm eq}(x_k) \cong 2\times 10^{-6}$.
 The analysis of ref.\
\cite{KT} shows that a good estimate of the relic abundance of 
$B_2$ is obtained by evaluating $Y_{B_1}$ (which is the source for 
$Y_{B_2}$ in the Boltzmann equation) at $x_f$: $Y_{B_2}(\infty) = 
Y_{B_1}(x_f)$.  On the other hand, the present abundance of stable
$B_2$ bosons must not exceed the observed DM abundance.  Using
baryons as a reference, 
\beq
	Y_{B_2}(\infty) <  
{\Omega_{DM} \over \Omega_b} {m_N\over \mu} \eta_b \cong 3\times 10^{-8}
\eeq
where $\Omega_{DM}/\Omega_b \cong 5$, $\eta_b\cong 6\times
10^{-10}$, $m_N$ is the mass of the nucleon, and we took $\mu=100$
MeV.  Putting these results together, we obtain the bound 
\beqa
	\epsilon &>& {1\over x_f}\left({3(\ln(\sfrac13\times 10^{8})-x_k +
\sfrac32\ln x_k))\over
	1.67\sqrt{g_*}\alpha}{\mu\over M_p}\right)^{1/2}\nonumber\\
	&=& {9\times 10^{-9}\over x_f}
\label{neweps}
\eeqa
Since $x_f< 28.7$ (the value when $\Delta E = 0$), this is approximately an order of
magnitude  stronger than
the bound (\ref{bbnc}) from nucleosynthesis.\footnote{If $x_f<x_k$,
then the bound is slightly modified since $B_1$ maintains equilibrium
density until $x_f$:
$$
		\epsilon > 
	{4\times 10^{-9}\over x_f}\left(17.3 + \sfrac32\ln x_f - x_f
	\right)^{1/2}
$$}

\subsubsection{Long-lived Higgs bosons}
\label{crdgb-hm}

We now consider the case where the Higgs boson $S$ that mixes with
the SM is the lightest metastable state of the dark sector.
The gauge bosons provide no more constraint in this case since they
are presumed to be heavier, and although they are stable, they
efficiently annihilate into dark sector Higgses with a negligible
relic density, $\sim (\mu/M_\chi)^2$ smaller than the closure
density.  

If the coupling of the Higgs to the SM is too strongly suppressed by
the small mixing angle $\theta$, there will be similar problem
as the one involving metastable gauge bosons, discussed above.
The Higgs should decay before nucleosynthesis to avoid dominating
the energy density of the universe.  We can directly adapt the result
(\ref{bbnc}) by replacing $\epsilon\to\theta$, $\mu\to m_S$, $e\to y_e$ (the electron
Yukawa coupling, $y_e = y_t m_e/m_t \cong 3\times 10^{-6}$):
\beq
	\theta > 4\times 10^{-6}\left({100{\rm\ MeV}\over
m_S}\right)^{1/2}
\label{thetabbn}
\eeq
Of course, this also forbids the possibility of a long-lived
intermediate state \cite{Rothstein} for transporting them outside the
galactic center before decaying into $e^+e^-$.  

\subsection{Long-lived intermediate DM states and direct detection
constraint}
\label{llis}

In section \ref{imh} we discussed the implications of an absolutely
stable intermediate DM state, protected by a discrete symmetry.  This
symmetry also made the models safe from downward transitions
$\chi_2\to\chi_1$ mediated by nuclear recoil in direct detection
experiments, since the gauge boson $B_3$ was forbidden from mixing
with the SM.  However, if the symmetry is not present and $B_3$ does
mix with hypercharge, interesting constraints can arise, since the
state $\chi_2$ generically has a lifetime longer than the age of the
universe, has a significant relic density, and can undergo
$\chi_2\to\chi_1$ in the detector \cite{BPR}.  The latter process is
not kinematically suppressed since it is exothermic, and it leads to
strong constraints on the mixing parameter $\epsilon$.  Ref.\
\cite{BPR} finds the 90\% c.l.\ limit $\epsilon<2\times 10^{-6}$ from
CDMS for $M_\chi=1$ TeV, $\delta M_{12}=100$ keV for the small
splitting which would be relevant for the iDM explanation of DAMA,
and $\mu = 1$ GeV.  As explained  above, we prefer $\mu = 100$ MeV,
which makes the constraint even more severe, 
\beq
	\epsilon<2\times 10^{-8}\left({\mu\over 100{\rm\
MeV}}\right)^2
\label{ddc}
\eeq
since the $\chi$-nucleon cross section scales like
$\epsilon^2/\mu^4$.

\begin{figure}[t] %\smallskip 
\centerline{\epsfxsize=0.4\textwidth\epsfbox{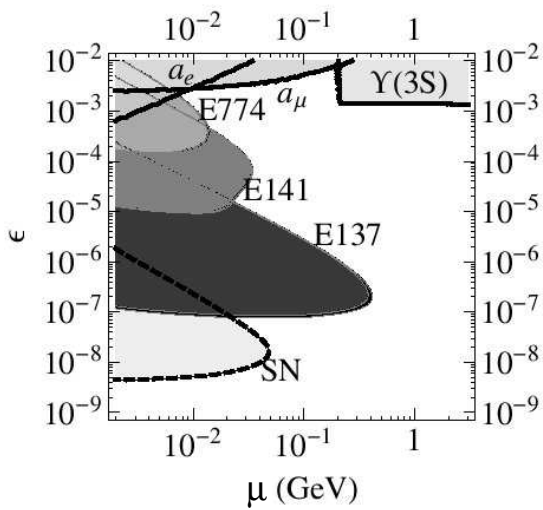}} 
\caption{Experimental constraints on $\epsilon$ versus gauge
boson mass $\mu$ taken from ref.\ \cite{Bjorken}.  Enclosed regions
are excluded by anomalous magnetic moments, beam dump experiments and
supernovae.} \label{eps-const}
\end{figure}

Notice that the window between  (\ref{ddc}) and our BBN or relic
density bounds
(\ref{bbnc},\ref{neweps}) is only a few orders of magnitude.  This region of 
parameter space is also below those which could be probed by
complementary experiments, as illustrated in fig.\ \ref{eps-const},
taken from ref.\ \cite{Bjorken} (see also ref.\
\cite{posp},\cite{batt}.)
In the models we
consider, the bound  (\ref{ddc}) can be evaded if we insist upon the
$Z_2$ symmetry which forbids the transitions leading to direct
detection.  This  makes it possible to have models which could
also be probed by laboratory experiments such as beam dumps.  Another
way to evade (\ref{ddc}) can arise if the mass splitting between the
intermediate and ground state is {\it too large} \cite{batt}, since
direct detection experiments do not look for very large recoil
energies.  The inverted mass hierarchy could thus be useful for this
purpose even if there is no $Z_2$ symmetry and the intermediate state
is only metastable.  

\section{Doublet dark matter}  
\label{ddm}

We now begin our investigation of more specific classes of models,
organized according to the SU(2) representation under which the DM
transforms.  
If the DM is in the doublet representation, 
it must be vector-like (Dirac) in order to have a bare mass term,
\beq
	M \bar\chi^i\chi_i
\eeq
In this case, DM number becomes conserved. 
Its abundance could be due to its
chemical potential rather than freeze-out, similar to the baryon
asymmetry, and so a nonthermal origin could be considered more
natural than for Majorana DM.

There is
no way to split the masses of the doublet through radiative
corrections from the gauge bosons, because each member of the doublet
has equal-strength interactions with all three gauge bosons.  For
example suppose only $B_1$ were to get a mass $\mu_1$; the contribution to
the $\chi$ mass matrix is $\delta M_{ik}=-\frac12\alpha\mu_1 
\tau^{1}_{ij} \tau^1_{jk}
= -\frac12\alpha\mu_1 \delta_{ik}$.  
But we
can get a splitting through the VEV of a triplet via the Yukawa
interaction
\beq
	h\chi^\dagger \tau^a \chi \Delta_3^a
\label{dyc}
\eeq
The suffix on $\Delta_3$ is a mnemonic for the fact that (for 
convenience) we 
take its VEV to be in the $a=3$ direction, since this gives the mass
splitting $\pm h\Delta_3$ between the Dirac states 
$\chi_1$ and $\chi_2$.

\subsection{Gauge kinetic mixing}
\label{gkmd}

Let us first consider the case of gauge kinetic mixing as the portal to the SM.
With the above mass splitting, either $B_1$ or $B_2$ must mix with the 
SM hypercharge so that transitions between $\chi_2$ and $\chi_1$ can occur,
with the production of $e^+ e^-$.  The triplet VEV which generates the
mass splitting is not suitable for generating the kinetic
mixing of the gauge boson via 
${1\over\Lambda} Y_{\mu\nu} B_a^{\mu\nu}\Delta_3^a$.  In fact, 
such mixing is dangerous from the standpoint of constraints from 
direct DM searches, since it would induce diagonal couplings via
$B_3$ of the  DM to nuclei.   One possibility is to have 
an additional triplet,
$\Delta_1^a$, coupling as in eq.\ (\ref{Bmixing3}),
which gets a VEV along the 1 (or 2) direction.  
The extra triplet VEV serves another purpose, by completely breaking
the SU(2) gauge symmetry, whereas a single triplet would break
SU(2)$\to$U(1).   Assuming that $\Delta_1$ gets its VEV along the
1 direction, the spectrum of the gauge bosons is
\beq
	\mu_1 =  g\Delta_3,\quad \mu_2 = g\sqrt{\Delta_3^2 +
\Delta_1^2},\quad \mu_3 = g\Delta_1
\eeq
With this spectrum and the couplings described above, 
$B_3$ is stable, but $B_2$ can decay into 
$B_3\,\gamma$ via the nonabelian gauge mixing interaction
$\epsilon g c_W F_{\mu\nu}B_2^\mu B_3^\nu$.   If we assume that
$\Delta_1>\Delta_3$, then annihilations $B_3 B_3\to B_1 B_1$
effectively deplete any potentially dangerous $B_3$ relic density
in the early universe.  

\begin{figure}[t] %\smallskip 
\centerline{\epsfxsize=0.45\textwidth\epsfbox{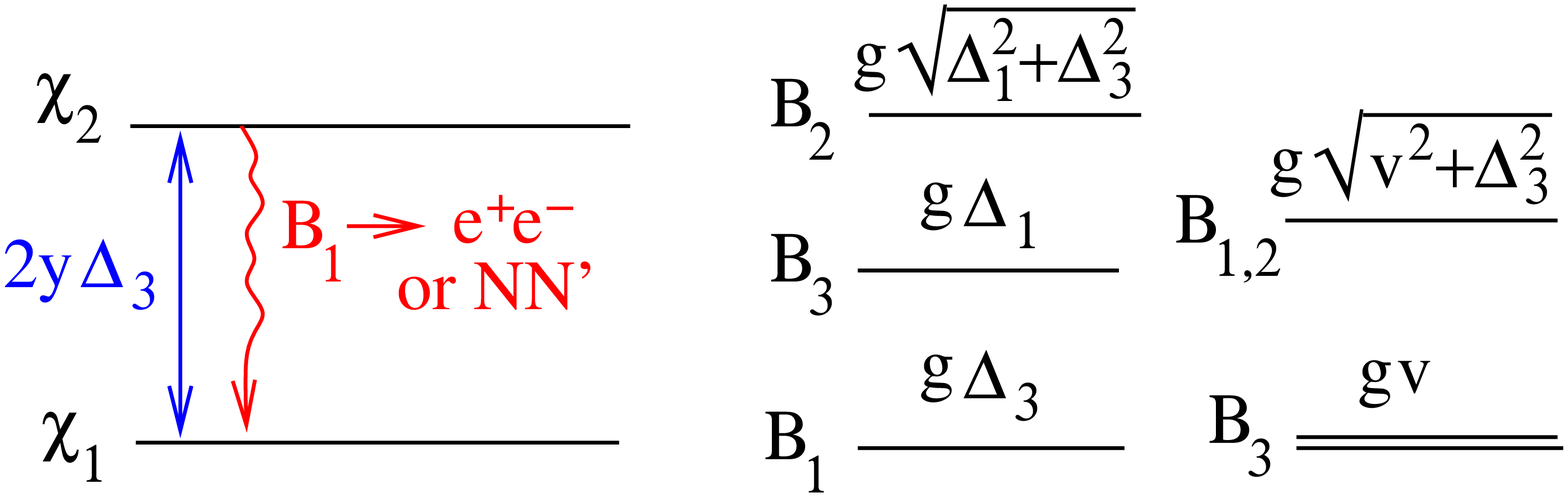}} 
\caption{Spectrum of doublet $\chi$ states (left) and two
possibilities for gauge boson spectra discussed in the text
(center and right).} 
\label{doublet}
\end{figure}

Alternatively, kinetic mixing could be accomplished by a Higgs doublet
as in  eq.\ (\ref{Bmixing2}), with VEV $h = (v/\sqrt{2})(1,1)^T$.  This would
cause mixing of only $B_1$ to the SM.  The gauge boson mass spectrum 
in this case is
\beq
	\mu_1 =  \mu_2 =  g\sqrt{\Delta_3^2 +
v^2}\quad  \mu_3 = g v.
\label{gbms}
\eeq
Similarly to the case of two triplets, $B_2$ can decay into 
$B_3\,\gamma$, but for this spectrum, the $B_3 B_3\to B_1 B_1$
annihilation channel is kinematically blocked.  Therefore the
mixing parameter $\epsilon$ must satisfy (\ref{neweps}) 
 to effectively deplete the relic $B_3$'s.
The mass levels for
the DM and gauge boson states are summarized in figure \ref{doublet}.

As mentioned in the previous section, doublet DM has the advantage of
allowing the gauge coupling to be as large as needed for getting the
right annihilation cross section, without additional constraints from
the size of the $\chi_1$-$\chi_2$ mass splitting, $\delta M =
2h\Delta_3$. For example one can adopt close to the preferred value
from section \ref{fits}, $\alpha_g=0.054$ (notice that this gives the
correct relic density for doublet DM, eq.\ (\ref{fra}), if
$M_\chi = 700$ GeV, compatible with the allowed region in
fig.\ \ref{meade-fig}), and take
the gauge boson mass at the 100 MeV scale, assuming the argument of
the previous section that diffuse gamma ray constraints prefer the
$4e$ annihilation channel over an admixture of $e$ and $\mu$.  The 
triplet VEV's are then of order $\mu/g \sim \mu\sim $ 100 MeV, and
the Yukawa coupling should be $h\sim 10^{-2}$ to accommodate the
excited DM (XDM) mechanism for explaining INTEGRAL/SPI.

\begin{figure}[t] %\smallskip 
\centerline{\epsfxsize=0.45\textwidth\epsfbox{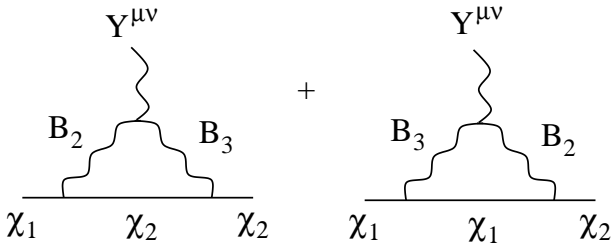}} 
\caption{Canceling loop diagrams contributing to the 
transition magnetic moment of doublet DM.} 
\label{doublet-loop}
\end{figure}

A distinctive feature of the doublet DM model is that its transition
magnetic moment is suppressed relative to that of triplet DM.  The
diagrams which contribute are shown in figure \ref{doublet-loop}.
The group theory factors from the DM gauge couplings of the two
diagrams are respectively $\tau^2_{22} \tau^2_{21}$ and $\tau^2_{21}
\tau^2_{11}$ (the Pauli matrices), which are equal and opposite.  Therefore the sum of the
diagrams is suppressed by the small mass difference $\delta M/M$ and
the gamma ray line from $\chi_2\to\chi_1\gamma$ decays is too
weak to be detected by INTEGRAL.   The constraint $\alpha_g < 0.08$,
eq.\ (\ref{alphacons}), does not apply.

\subsection{Higgs mixing}

Since we have argued that a DM Yukawa coupling (\ref{dyc}) to a
triplet is already necessary to get the doublet mass splitting, it is
tempting to make a more economical model without gauge kinetic
mixing, by letting this triplet mix with the SM Higgs through a
$\lambda |H|^2 |\Delta_3|^2$ coupling. The most stringent of the 
constraints on $\lambda$  from section \ref{cdc} is  (\ref{ddc2}), 
arising from direct detection of the DM.  For $h=10^{-2}$ this gives
$\lambda < 3\times 10^{-2}$ if  $m_{\Delta_3}\cong 200$ MeV and
$\langle\Delta_3\rangle\cong 1$ GeV.  Saturating this inequality
leads to the mixing angle $\theta\cong 2\times 10^{-4}$, according
to (\ref{ddc1}).   This value is consistent with our BBN constraint
(\ref{thetabbn}).

\subsection{Diagonal couplings to $B_3$}

Even though we took care to avoid the direct annihilation channel
$\chi_1\chi_1\to e^+ e^-$ through virtual $B_3$ production, by
forbidding mixing between $B_3$ and the SM hypercharge, it is
impossible to forbid $\chi_1\chi_1\to B_1 B_2$.  If only $B_1$ 
 couples to the SM but not $B_2$, this results in the final
state $e^+ e^- B_2$, where $B_2$ is invisible.  In the foregoing
we have noted that the two-body final state $e^+ e^-$ is ruled out,
because its spectrum has the wrong shape to fit the PAMELA and Fermi
observations.  The three-body final state is much more similar to
the four-body one in this respect, however, because the two visible
leptons share the energy of the incoming $\chi$'s with the $B_2$.
They thus have a soft spectrum which is qualitatively similar to that
of the four-body case.  Thus the $\chi\chi\to e^+ e^- B_2$ channel
in this model is on a similar footing to the $\chi\chi\to 4e$ one
in models with Majorana DM.  If $B_2$ also mixes with the SM
hypercharge so that $B_2\to e^+ e^-$,  they become identical.

\section{Triplet Dark Matter}  
\label{tdm}
We now take $\chi_a$ to be a
real (Majorana) triplet of SU(2).  It can have a bare Majorana mass
$M\chi^a\chi^a$.  
In this case, mass splittings can be generated
radiatively, as well as at tree level.  A doublet VEV $h$ gives
equal contributions to all the gauge boson masses, so it does not
generate any mass splittings between the $\chi_a$'s.  It is thus more
economical to  assume there is at least one triplet VEV contributing
to the SU(2) breaking.  However
this is not enough to fully split the DM states, since a single
triplet VEV would leave two of the gauge bosons degenerate in mass,
and the radiative corrections would then do likewise for the DM
states.  We are led to  introduce either a second Higgs triplet as in the
doublet DM case, or a quintuplet.  It is also interesting to consider a
model with three  triplet VEVs, since this gives additional
freedom in arranging the DM spectrum to have an inverted or normal
hierarchy.\footnote{We prefer the inverted hierarchy since it can boost
the effectiveness of XDM, and avoid the direct detection constraint
(\ref{ddc}).}
In the following, we consider these different Higgs sectors and the 
$\chi^a$ mass splittings that arise due to radiative corrections.  

\subsection{Two triplet Higgs fields}
Let us turn on VEV's for two triplets in orthogonal directions,
$\Delta_1^1$ and $\Delta_2^2$, for example.  It is easy to write a
Higgs potential whose minima have this property:
\beq
	V = \sum_i \lambda_i(\Delta_i^2 - v_i^2)^2 + \lambda_{12}
	(\vec\Delta_1\cdot\vec\Delta_2)^2
\label{pot2}
\eeq
As long as $\lambda_{12}>0$, the energy is minimized for orthogonal
VEVs.

\subsubsection{Mass spectra}

\begin{figure}[t] %\smallskip 
\centerline{\epsfxsize=0.35\textwidth\epsfbox{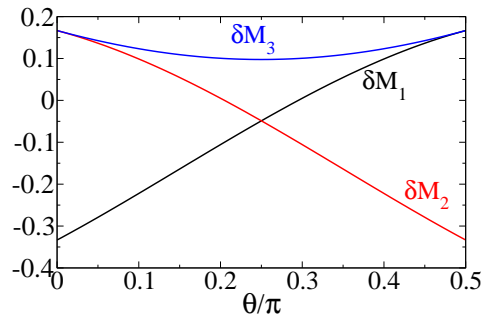}} 
\caption{Mass splittings of triplet $\chi$ states for the model with two
orthogonal triplet VEVs as a function of $\theta =
\tan^{-1}(\Delta_2/\Delta_1)$, eq.\ (\ref{3spect}), in units of 
$\alpha g \sqrt{\Delta_1^2+\Delta_2^2}$.} 
\label{spect3}
\end{figure}

The gauge boson mass
spectrum is
\beq
	\mu_1 = g\Delta_2,\quad \mu_2 = g\Delta_1,\quad \mu_3 =
	g\sqrt{\Delta_1^2 + \Delta_2^2}
\label{3gbspect}
\eeq 
(With no loss of generality, one can take $\Delta_1^1,\Delta_2^2>0$
by doing a global gauge transformation.)
The radiative corrections to the DM masses are
\beqa
	\delta M_1 &=& -\frac12 g\alpha 
	\left(\Delta_1 + \sqrt{\Delta_1^2 + \Delta_2^2}\right)\nonumber\\
	\delta M_2 &=& -\frac12 g\alpha
	 \left(\Delta_2 + \sqrt{\Delta_1^2 + \Delta_2^2}\right)\nonumber\\
	\delta M_3 &=& -\frac12 g\alpha (\Delta_1 + \Delta_2)
\eeqa
Depending on the ratio $\Delta_2/\Delta_1$, this can correspond to
either the normal or inverted hierarchy.  To see the range of
possibilities, define $\Delta_1 = \Delta\cos\theta$ and 
$\Delta_2 = \Delta\sin\theta$, and subtract from each $\delta M_i$
the average splitting (since this just renormalizes the bare value
$M_\chi$):
\beqa
	\delta M_1 &\to & -\frac16 g\alpha \Delta
	\left(1+\cos\theta-2\sin\theta\right)\nonumber\\
	\delta M_2 &\to& -\frac16 g\alpha\Delta
	 \left( 1+\sin\theta-2\cos\theta\right)\nonumber\\
	\delta M_3 &\to & -\frac16 g\alpha\Delta \left(-2 +
	\cos\theta+\sin\theta\right)
\label{3spect}
\eeqa
The spectrum is plotted as a function of $\theta$ (which lies in the
range $[0,\pi/2]$ due to our requirement that both VEVs be positive)
in figure \ref{spect3}.  There it is clear that the inverted hierarchy
occurs if $\theta\ll 1$ ($\Delta_2\ll\Delta_1$) or $\theta \cong \pi/2$
($\Delta_1\ll\Delta_2$), while the normal one occurs if
$\theta\cong\pi/4$ ($\Delta_1\cong\Delta_2$).

\subsubsection{Inverted hierarchy and $Z_2$ symmetry}

To discuss the phenomenology of this model, we must specify which
of the gauge kinetic mixing operators $\vec\Delta_i\cdot \vec
B_{\mu\nu}F^{\mu\nu}/\Lambda_i$ are assumed to be turned on.   The simplest
possibility, and the one that allows for $Z_2$ symmetry, is that
only one of them is significant, say the one corresponding to $\vec
\Delta_1$.  Then only $B_1$ mixes with the SM, and we can assign
$Z_2$ charges to $B_2$, $B_3$, $\chi_2$, $\chi_3$.  The uncharged
state $\chi_1$ cannot decay into $\chi_2$, so to implement the
inverted hierarchy for INTEGRAL, we should choose $\theta\lsim \pi/2$
to make $\chi_1$ the intermediate state.   

By choosing $\theta\cong\pi/2$, hence $\Delta_1\ll\Delta_2$, we
obtain from (\ref{3gbspect}) the gauge boson mass spectrum $\mu_2
<\mu_1\lsim \mu_3$.  According to the argument of section
\ref{crdgb-gkm},
the gauge mixing parameter must then exceed the lower bound
(\ref{neweps}).

To be compatible with a nonthermal origin, the gauge coupling must
be larger than $\alpha_g=0.03$, according to eq.\ (\ref{fra}).
Taking  $\alpha_g=0.06$ and 
 $\mu\cong g\Delta =100$ MeV, for example,    the
gauge coupling is then $g=\sqrt{4\pi\alpha_g} = 0.87$, and $\Delta =
\mu/g = 115$ MeV.    The largest mass splitting is of order $\frac12
\alpha\mu = 3$ MeV.  We have the freedom to adjust the smaller
splitting as desired by choosing 
$\theta =\tan^{-1}(\Delta_2/\Delta_1)$.  Taking $\theta=0.4\pi$
gives $\delta M_{23} = 2.1$ MeV and $\delta M_{13} = 150$ keV, which
is small enough to comfortably enhance the 511 keV signal to the
level observed by INTEGRAL \cite{new}.

\subsubsection{Normal hierarchy}

Since it might be argued that the window for iDM to explain the
DAMA/LIBRA annual modulation  is not completely closed \cite{crucis},
for completeness we consider the case of the normal mass hierarchy.
As is clear from fig.\ \ref{spect3}, it arises from
choosing $\theta$ close to $\pi/4$.   If one wants to have both the
iDM and XDM effects for DAMA and INTEGRAL, respectively, then $B_3$
must mix with the SM hypercharge, in addition to $B_1$ (if
$\theta>\pi/4$) or $B_2$ (if $\theta<\pi/4$).  Interestingly, the
spectrum (\ref{gbms}) shows that the boson which does not mix with
the SM is the lightest one, while $B_3$ is the heaviest.  Therefore
the lightest gauge boson is stable and the BBN bound (\ref{neweps})
on $\epsilon$  applies to this model.

With the normal hierarchy there is no requirement for a nonthermal
origin of the DM, so we consider the value $\alpha_g=0.03$, eq.\
(\ref{fra}), needed for the correct thermal relic density,
and the boost
factor 300 needed for PAMELA/Fermi.  The gauge coupling is $g=0.61$.  Fig.\ 1 of ref.\
\cite{AH} shows that this value of the coupling gives 
approximately the required value of the
boost factor for $\mu \lsim 1$ GeV, assuming the DM velocity dispersion
of $\sigma = 150$ km/s.  We are free to adjust the
triplet VEVs to obtain the desired mass splittings.  For example
with $\delta M_{13}\sim\delta M_{23} \cong 2$ MeV, one finds
$\Delta_1\sim\Delta_2\cong 750$ MeV.  To get a small mass splitting
$\delta M_{12} = \frac12 g\alpha_g(\Delta_1-\Delta_2)$ of order
100 keV, if one wishes to explain DAMA, the two VEV's $\Delta_1$
and $\Delta_2$ have to be tuned to be equal to each other to within
one part in 70.

\subsubsection{Nonorthogonal VEVs}

For a generic Higgs potential, $\Delta_1$ and $\Delta_2$ are not
orthogonal (\textit{e.g.}, $\Delta_2$ could be nonzero in both the
$1$ and $2$ directions.). Aside from changing the details of the
gauge masses and fermion mass  splittings, this has the same effect
as turning on kinetic mixing terms for both $B_1$ and $B_2$.  This is
because  the mass eigenstates of the vectors  become mixtures of
these two directions; the gauge interactions of the mass  eigenstates
with the $\chi_a$ can be put in canonical form with a  corresponding
rotation of $\chi_{1,2}$.  In this basis, the $\chi_a$ are mass
eigenstates.  However, the $B$ vector
that mixes with the SM hypercharge vector is a linear combination
of both the $B_1$ and $B_2$ mass eigenstates, 
so both the $\chi_2\lra
\chi_3$ and $\chi_1\lra\chi_3$ transitions couple to the SM.  This
implies there is no $Z_2$ symmetry protecting the intermediate
state, in the inverted hierarchy case.  Thus it is important to keep
the VEVs orthogonal in that case, whereas relaxing this assumption
does not hurt the normal hierarchy scenario.  In fact it makes it
simpler, by requiring only a single gauge kinetic mixing term to be
nonnegligible, while still coupling both DM transitions to the 
SM, as required by the iDM and XDM mechanisms.

%the additional gauge kinetic mixing only provides a means to
%depopulate the intermediate DM state by scattering off SM particles. 
%We note that $\chi^3$ is always the  heaviest DM state.

\medskip
\subsection{Three triplet Higgs fields}

\subsubsection{Generic VEVs}

For a generic Higgs potential with three triplet Higgs fields, we can
use gauge transformations to align $\vec\Delta_1$ in the 1 direction,
with $\Delta_1^1>0$, and $\vec\Delta_2$ in the 1-2 plane with 
$\Delta_2^2>0$, while the direction of  $\vec\Delta_3$  remains
general.   The vacuum manifold can thus be parametrized  by an
overall amplitude $\Delta = (\sum_i |\vec\Delta_i|^2)^{1/2}$ and five
angles: two measuring the relative amplitudes of the three VEVs 
(covering 1/8 of a sphere), one determining the orientation of
$\vec\Delta_2$ (covering 1/2 of a circle), and two controlling the
orientation of $\vec\Delta_3$ (on a full sphere).  Depending on the
vacuum state, the DM mass splittings can take on any hierarchy. 
Unlike the previous case, $\chi_3$ need not be the heaviest DM state.

At a generic position in the vacuum space, all the $B$ mass
eigenstates mix with the SM hypercharge vector. Assuming $m_1\lesssim
m_2 \ll m_3$ in a  normal hierarchy, one can implement both the iDM
and XDM dark matter excitation mechanisms.  There are transition
magnetic moments between all pairs of $\chi_i$ and $\chi_j$ allowing
for single-photon decays of $\chi_2$ and $\chi_3$.  The exception is
for  $\Delta_2^1=\Delta_3^1=0$; in that case, there is an unbroken
$Z_2$ symmetry, and only $B_1$ mixes with the SM.  Then either iDM or
XDM is possible (not  both), and one of the excited DM states will be
stable.

%we now have the transition $\chi_1 N\to\chi_2 N'$ for 
%the iDM explanation of DAMA.  With a favorable DM velocity distribution, 
%this model can also provide the XDM effect for the INTEGRAL/SPI 511 keV line.
%However, for generic VEVs, no symmetry protects the excited DM states from
%decaying.  Therefore, it is possible for $\chi^{2,3}$ to decay to 
%$\chi^1$ by multi-photon emission.  The exception is for 
%$\Delta_2^1=\Delta_3^1=0$; in that case, there is an unbroken $\Z_2$ symmetry,
%and only $B^1$ mixes with the SM.  Then either iDM or XDM is viable (not 
%both), but one of the excited DM states may be stable.

\subsubsection{Orthogonal VEVs}
\label{ortho}

The potential (\ref{pot2}) can be generalized to one that
leads to three orthogonal
triplet VEVs:
\beq
	V = \sum_i \lambda_i (\Delta_i^2 - v_i^2)^2 + 
	\left[\eta_1 (\vec\Delta_2\cdot\vec\Delta_3)^2 + \hbox{cyc.\ perm.}
	\right]
\label{pot3}
\eeq
As long as the $\eta_i$ couplings are positive, the desired vacuum
state is a minimum of the potential.  There is no obstacle to
assuming that  only $B_1$ mixes with the
SM vectors, if one wants to incorporate the 
 $Z_2$ symmetry that prevents $\chi_1$ from decaying.
We will show that both normal and inverted DM mass hierarchies are
possible, with $\chi_1$ as the intermediate state. 
Many configurations of the VEVs are compatible with the 
iDM and/or XDM mechanisms.  

%In this case, $\chi^1$ is essentially a spectator, since
%it does not couple to $B_1$.
%JC: unless we turn on mixing for B_2 or B_3.

\begin{figure}[t] %\smallskip 
\centerline{\epsfxsize=0.5\textwidth\epsfbox{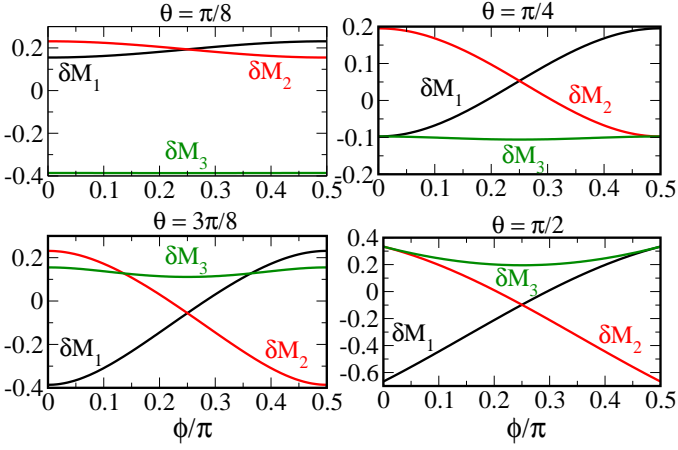}} 
\caption{Spectrum of triplet $\chi$ states for model with three
triplet VEVs parametrized as in (\ref{dangles}) and with the 
average mass shift subtracted.  $\delta M_i$ are in units of
$g\alpha_g\Delta/2$.
Each panel shows a different value of $\theta$.} 
\label{spect33}
\end{figure}

Taking each of the respective fields $\vec\Delta_i$ to align along
the $i$th direction, the gauge boson masses are given by
$\mu_1 = g\sqrt{\Delta_2^2 + \Delta_3^2}$ and cyclic permutations,
while the $\chi$ mass splittings are $\delta M_1 = -\frac12 
\alpha_g \left(\mu_2+\mu_3\right)$ plus cyclic permutations.
The full range of possibilities for the spectrum can be explored by
parametrizing the VEVs in spherical coordinates,
\beqa
	\Delta_1 &=& \Delta\sin(\theta)\cos(\phi)\nonumber\\
	\Delta_2 &=& \Delta\sin(\theta)\sin(\phi)\nonumber\\
	\Delta_3 &=& \Delta\cos(\theta)
\label{dangles}
\eeqa
where $\Delta=(\sum_i\Delta_i^2)^{1/2}$, and the angles are restricted
by $0\le\theta\le\pi/2$ and $0\le\phi\le\pi/2$ so that
each $\Delta_i$ is positive.  The resulting $\delta M_i$'s, 
shifted to set the average value $\frac13\sum_i \delta M_i$ to zero,
are shown in figure \ref{spect33}.

To obtain the inverted hierarchy, where  $\chi_1$ is the 
intermediate state with a mass close to the heaviest state,  one
possibility is to take  $\Delta_3\ll \Delta_1,\Delta_2$, which
essentially reproduces the two Higgs case studied above.  This 
corresponds to $\theta\cong\pi/2$  and $\phi\lsim \pi/2$
in fig.\ \ref{spect33}, with $\chi_2$ and $\chi_3$ being respectively
the ground state and highest excited state.  The other possibility is
to take $\Delta_3\gg\Delta_1\gtrsim\Delta_2$, corresponding to
$\theta\cong 0$ and $\phi<\pi/4$, in which case $\chi_3$ is the
lowest mass state.    From fig.\ \ref{spect33} one can also see
examples of the normal hierarchy, for example near $\theta=\pi/2$
and $\phi=\pi/4$.

It is interesting to notice that smaller mass
splittings than the generic scale $\alpha_g\mu$ can be obtained
 near special values of the VEVs.
When $\Delta_1$ and $\Delta_2$ are equal, $\phi=\pi/4$,
the masses $M_1$ and $M_2$ become
accidentally degnerate.  By tuning the VEVs to be close to this
point, the 100 keV scale desired for the iDM splitting can be achieved
even if $\alpha_g\mu$ has the right magnitude for getting the XDM
splitting.

\subsection{Quintuplet Higgs field}
\label{tqh}

 Allowing DM to couple to
a quintuplet Higgs field, which is a symmetric traceless tensor
$\Sigma_{ab}$, gives further flexibility in model building, since the
pattern of gauge boson masses induced by the $\Sigma$ VEV is different
than for triplets, and one also has the possibility of a Yukawa
coupling  $y \chi_a \Sigma_{ab}\chi_b$
to give tree-level contributions to the DM mass splitting.  In
addition one still
wants at least one triplet or doublet Higgs to generate kinetic mixing
of one of the $B$'s to the photon.  

\begin{figure}[t] %\smallskip 
\centerline{\epsfxsize=0.4\textwidth\epsfbox{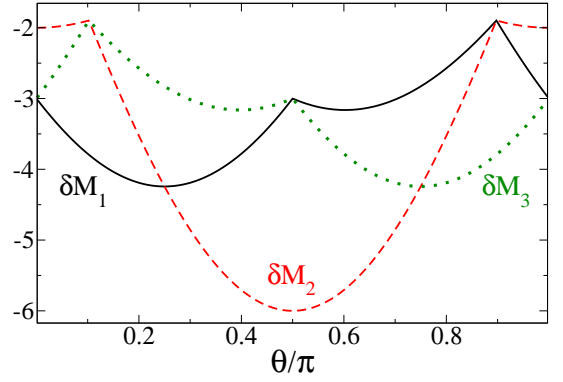}} 
\caption{Mass splittings of triplet $\chi$ states for model
with two components of quintuplet VEV, eq.\ (\ref{qvev}), 
parametrized by $\tan\theta = B/A$.  
$\delta M_i$ are in units of
$\frac12 g\alpha_g\sqrt{A^2+B^2}$.} 
\label{spect35}
\end{figure}

For a fairly general class of potentials,  
the VEV of $\Sigma$ can be chosen to be along the diagonal components.  
Let us take this as a simplifying assumption and show how much can be
accomplished with just the two components, which we denote by
\beq
	\langle \Sigma\rangle = {1\over\sqrt{2}}\left(\begin{array}{ccc}
	A -B& & \\
	 & 2B & \\
	 &   & -A-B\end{array}\right)
\label{qvev}
\eeq
(note the different normalization of $B$ than for the corresponding
quintuplet $\chi$ field in (\ref{5plet})).
If $\langle \Sigma\rangle$ is much larger than the triplet or doublet
VEV, the resulting spectrum of gauge boson masses is approximately
\beq
	\mu_1 = g \left|A + 3B\right|,\quad
	\mu_2 = g\, |2A |,\quad
	\mu_3 =  g \left|A - 3 B\right|
\eeq
where we used the generators
\beq
 T^d_{ab,ce} \sim
i(\epsilon_{adc}\delta_{be} + \epsilon_{bde}\delta_{ac})
\label{5gen}
\eeq 
where ``$\sim$'' denotes that the expression must be symmetrized on $ab$ and $ce$.
The radiative mass corrections  of the DM states are
\beqa
	\delta M_3 &=& -\frac12 g\alpha_g \left( \left|A+3
	B\right| + |2A|\right)\nonumber\\
	\delta M_2 &=& -\frac12 g\alpha_g \left( \left|A+3
	B\right| + \left|A-3
	B\right|\right)\nonumber\\
	\delta M_1 &=& -\frac12 g\alpha_g \left( \left|A-3
	B\right| + |2A|\right)
\label{t5spect}
\eeqa
Parametrizing the VEVs by $\tan\theta = B/A$, we obtain the full range
of possibilities by letting $\theta$ range from 0 to $\pi$.  Although
the region $\pi/2 < \theta < \pi$ can be mapped onto $0<\theta<\pi/2$
by gauge transformations, the freedom to do so is generally inhibited
by VEVs of other Higgs fields such as triplets, which we mention
below.  The result is shown in fig.\ \ref{spect35}.

\subsubsection{Normal hierarchy}

Fig.\ \ref{spect35} shows that the normal hierarchy occurs in the
region of $\theta\cong 0$ and $\theta\cong\pi/4$ (with the same shape
of spectra at $\theta\cong\pi$ and $\theta\cong 3\pi/4$). For
illustration consider the case of $\theta\cong 0$, which corresponds
to $B\ll A$.  The gauge bosons have the spectrum
$\mu_3\lsim\mu_1<\mu_2$. The order of the DM masses is $M_1\lsim
M_3<M_2$, and we can turn on transitions between $\chi_1$-$\chi_2$
and $\chi_1$-$\chi_3$ which couple to the electron vector current, by
mixing $B_3$ and $B_2$ with SM hypercharge.   This can be
accomplished using two triplets, which we will call $\vec\Delta_2$
and $\vec\Delta_3$,  with VEV's $\Delta_2^2$ and $\Delta_3^3$.  A
model-building  challenge is to find a scalar potential which gives
rise to this symmetry breaking pattern together with that assumed for
the quintuplet.  

We noted above that it is easy to make a potential
for triplets that gives rise to orthogonal VEV's.  Suppose we do this;
then global SU(2) transformations can be used to orient them in the
2 and 3 directions, respectively.  Next consider the $\Sigma$ sector.
The term $\lambda({\rm tr}\Sigma^2 - v^2)^2$ is O(5) symmetric under
rotations of the vector $(A,B/\sqrt{3},C,D,F)$, where $C,D,F$ are the
off-diagonal components of $\Sigma$.  To break this symmetry in such
a way as to prefer the $A,B$ components, we can add terms
\beq
	\Lambda_2 \Delta_2^T \Sigma \Delta_2 
	+ \Lambda_3 \Delta_3^T \Sigma \Delta_3
\label{break}
\eeq  
which are linear in $A$ and $B$ when the $\Delta_i$ get their expected
VEVs, and thus lead to nonzero VEVs for $A$ and $B$.  This would
be spoiled by a term of the form $\Delta_2^T \Sigma \Delta_3$, but
the latter can be forbidden by separate discrete symmetries under
which $\Delta_2$ or $\Delta_3$ change sign.  These symmetries are
weakly broken by the gauge kinetic mixing terms, which would
presumably give rise to a small $\Delta_2^T \Sigma \Delta_3$ interaction
through loops.  This would generate perturbations to the 
previous analysis due to the presence of small off-diagonal VEVs
in $\Sigma_{ab}$.  

\subsubsection{Inverted hierarchy}

Fig.\ \ref{spect35} also reveals the inverted hierarchy at
$\theta\cong 0.1\pi,$ $0.9\pi$ and $\pi/2\pm\epsilon$.  The latter occurs
when $|A|\ll |B|$.  Consider the case $\pi/2-\epsilon$ where
$M_2 < M_1 \lsim M_3$.  We need $B_1$ to mix with the SM in this case,
suggesting a triplet $\vec \Delta_1$ with VEV in the component
$\Delta_1^1$.  One can use an analogous potential to (\ref{break}),
$\Lambda_1 \Delta_1^T \Sigma \Delta_1 + \Lambda_2 \Delta_2^T \Sigma 
\Delta_2$, to generate VEVs in the $A,B$ components, if we add the
additional triplet $\vec \Delta_2$, which however does not play any
role in the gauge kinetic mixing.

For this scenario, the gauge bosons have the spectrum
$\mu_2 < \mu_3 < \mu_1$.  Thus $B_1$ which mixes with the SM is
the heaviest.  The relic gauge boson  constraint (\ref{neweps})
then applies, making this model susceptible to laboratory searches
for light gauge bosons that mix with the SM.  

\section{Quintuplet dark matter}
\label{qdm}

As the highest DM representation we will consider here, we turn to
the quintuplet case, where $\chi_{ab}$ is a traceless symmetric
tensor.  The gauge generators in this representation are given
in (\ref{5gen}), which for conciseness is not symmetrized in its
indices, but the actual generator must be symmetrized in
$ab$ and $ce$, with accompanying factor of $1/4$.  We will label the canonically normalized 
states of $\chi_{ab}$ by
\beq
	\langle \chi\rangle = \left(\begin{array}{ccc}
	A-B/\sqrt{3}& C  & D\\
	C  & 2B/\sqrt{3} & F\\
	D  & F  & -A-B/\sqrt{3}\end{array}\right)
\label{chiab}
\eeq
(Notice the change in normalization of $B$ compared to our choice
for quintuplet Higgs fields in (\ref{qvev}).)\ \ 
These are the mass eigenstates at tree level.

\subsection{Radiative mass corrections}

\begin{figure}[t] %\smallskip 
\centerline{\epsfxsize=0.45\textwidth\epsfbox{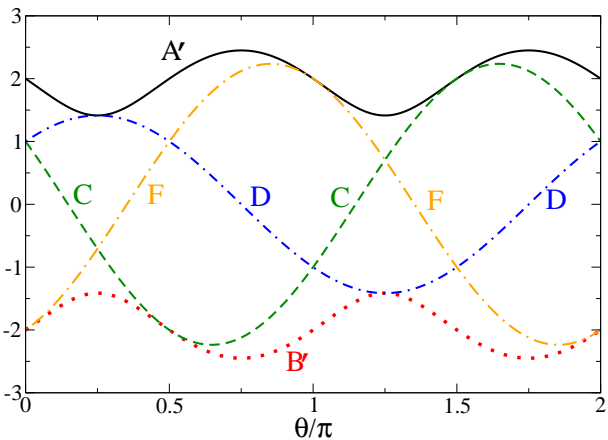}} 
\caption{Spectrum of quintuplet states as a function of $\theta$
which parametrizes gauge boson splittings.  Units of $\delta M_i$
are $2\sqrt{2}\alpha[(\mu_1-\mu_2)^2 + (\mu_3-\mu_2)^2]^{1/2}$.} \label{5spect}
\end{figure}

In previous sections we have given explicit expressions for the
gauge boson masses assuming various patterns of symmetry breaking.
Here we will leave them unspecified and study the $\chi$ radiative
mass splittings as a function of general values of $\mu_i$.  
The $\chi$ mass splitting term is given by
\beq
\delta{V}_{\rm mass} =	-\frac12\alpha\sum_d\ \mu_d\, 
	\chi_{\lower 0.25em \hbox{$\scriptstyle ab$}}\, 
	T_{ab,ce}^d T_{ce,fg}^d\, 
	\chi_{\lower 0.25em \hbox{$\scriptstyle fg$}}
\eeq
For general values of the gauge boson masses, we find
\beqa
	\delta{V}_{\rm mass} = &&-\alpha \mu_1((A+\sqrt{3}B)^2+ C^2+ D^2 + 4F^2)\nonumber\\
	&&-\alpha\mu_2(4A^2 +C^2 +4D^2 + F^2)\\
	&&-\alpha \mu_3((A-\sqrt{3}B)^2+ 4C^2+ D^2 + F^2)\nonumber
\eeqa
In the simpler case where $\mu_1 = \mu_3$, 
the terms which mix $A$ and $B$ cancel and the mass terms are 
diagonal.  The average mass splitting in this case 
is $-4\mu_1-2\mu_2$.  Subtracting away this central value, we obtain
the hierarchy of mass splittings
\beq
(A,D): +2\delta,\quad (C,F): -\delta,\quad B: -2\delta
\eeq
where $\delta = 4\alpha_g(\mu_1-\mu_2)$.    In the more general case
where $\mu_1=\mu+\delta_1/4\alpha_g$, $\mu_2=\mu$, $\mu_3 = 
\mu+\delta_3/4\alpha_g$,
we get splittings equal to
\beqa
	&&(A',B'): \pm 2 \sqrt{\delta_1^2-\delta_1\delta_3+\delta_3^2}\nonumber\\
	&&C: \delta_1-2\delta_3,\quad D: \delta_1+\delta_3,\quad
	F:\delta_3-2\delta_1
\eeqa
These can be parametrized using
\beq
	\delta_1 = \delta\cos\theta,\quad \delta_3 = \delta\sin\theta
\eeq
so that $\delta^2\equiv 8\alpha_g^2[(\mu_1-\mu_2)^2 
+ (\mu_3-\mu_2)^2]$ controls the overall magnitude of the splittings, whereas
$\theta$ controls the relative values (and can be in the range
$[0,2\pi]$ since there is no restriction on the signs of the gauge
boson mass differences).  We get
\beqa
	\delta M_{A'} &=& +2\delta\sqrt{1-\sfrac12\sin(2\theta)} \nonumber\\     
	\delta M_{B'} &=& -2\delta\sqrt{1-\sfrac12\sin(2\theta)}\nonumber\\ 
      \delta M_C &=& \delta\left[\cos(\theta)-2\sin(\theta)\right]\nonumber\\ 
      \delta M_D &=& \delta\left[\cos(\theta)+\sin(\theta)\right]\nonumber\\ 
      \delta M_F &=& \delta\left[\sin(\theta)-2\cos(\theta)\right]
\eeqa
The spectrum as a function of
$\theta$ is shown in fig.\ \ref{5spect}.  The complete range of
possibilities for splitting by radiative corrections alone is thus
encompassed in the figure, assuming given values of the $\mu_i$ can
be achieved by the appropriate choice of scalar VEV's.  

\subsection{Gauge interactions and mass hierarchies}

The interactions of the gauge bosons with the quintuplet states are
off-diagonal, as expected for Majorana particles.  Suppressing Lorentz
indices and gamma matrices, they are
proportional to
\beqa
	{\cal L}_{\rm gauge} &\sim&B_1(DC+\sqrt{3}FB + FA)+
B_2(2AD-FC)\nonumber\\ &&+ B_3(CA+\sqrt{3}BC +
FD)
\label{qgbi}
\eeqa
The transitions which can be mediated are shown in figure
\ref{5trans}.  This diagram is useful for determining what kinds of
DM mass spectra can be consistent with explaining the various
experimental observations.

\subsubsection{Inverted hierarchy and $Z_2$ symmetry}
\label{ihz2}

Here we give some examples of the inverted mass hierarchy 
with $Z_2$ symmetry which can help boost the production of 
low-energy positrons as observed by INTEGRAL/SPI. 
In section \ref{z2fq} we identified the possible discrete
symmetries which could exist for a given choice of gauge
kinetic mixing.  Consider the case where $B_3$ mixes with 
hypercharge.  According to the arguments in section \ref{z2fq},
either $D$ or $F$ should be chosen as the intermediate state.
Fig \ref{5spect} shows that at $\theta\lsim \pi/4$
$F$ can be the stable intermediate state with a small mass gap
below $C$.  Scattering processes $FF\to CC$ in the galaxy can be
enhanced, followed by $C\to B'e^+e^-$ via $B_3$ exchange.
Similarly, at $\theta \gsim \pi$, $D$ can be chosen as the
intermediate state, giving rise to $DD\to CC$ followed by 
$C\to B'e^+e^-$.  Both of these examples have analogous counterparts
just on the other side of the degeneracy between states.
At $\theta\gsim \pi/4$, the roles of $C,F$ and $B_1,B_3$ 
are interchanged, while for $\theta \lsim \pi$, the roles of
$C,D$ and $B_2,B_3$ are interchanged.

From fig.\ \ref{5spect} we identify several other possible
inverted hierarchy realizations: $\theta\cong 0$, with $C,D$ as the
topmost relevant states,  $\theta\cong\pi/2$, involving $D,F$ and
$\theta\cong 5\pi/8$, involving $C,F$, and $\theta\cong 3\pi/2$ with
$D,F$.  In short, near every place where two mass eigenvalues cross
at an angle, one can have an inverted mass hierarchy.  There are
six such values of $\theta$ where this occurs.

\subsubsection{Normal hierarchy}

If one prefers a model with normal mass hierarchy, fig.\ \ref{5spect}
shows that there are several possibilities, close to points where the
$B'$ curve is tangent to that of $C,D$ or $F$.  These occur at
$\theta=0,$ $\pi/2$, $5\pi/4$.    Notice
that very small mass splittings can be arranged near these points
with relatively little tuning of $\theta$ due to the fact that
the curves are tangent to each other.  This gives another
way of obtaining smaller splittings than the generic size.

Curiously, in no case can the
heaviest state $A'$ be relevant for XDM, because there is no
gauge interaction which couples it to the lightest state $B'$. 
Instead $A'$ is a spectator, and the highest relevant
state is either $C,D,F$, one of  which happens to be degenerate with
$A'$ at the angles $\theta=\pi/4$, $\pi$ or $3\pi/2$. 

\subsubsection{No combined hierarchy}

Because of the extra complexity of the quintuplet spectrum and
gauge couplings, it is tempting to look for a situation where both
the normal and the inverted hierarchies could exist simultaneously,
combining the advantages of the latter for XDM while still
leaving open an iDM explanation for DAMA.   This turns out to be
impossible, however.  First consider the situation where only one
gauge boson mixes with the SM, say $B_1$.    To be compatible with both iDM and
XDM, $B_1$ would have to mediate transitions between
the ground state $B'$, an admixture  of $A$ and $B$, and two
other states.  Perusal of the transitions in figure \ref{5trans}
shows that no gauge boson has this property.  

The next alternative is that there are two gauge bosons which mix with
the SM, one for the iDM transition and one for the XDM.  The problem
here is that then $Z_2$ symmetry is broken for two gauge bosons.
The $B_1 B_2 B_3$ gauge interaction then forces it to be broken for
all of them, and no $Z_2$ exists to protect the higher intermediate
state for XDM from decaying.  

\section{An SU(2)$\times$U(1) model} 
\label{su2u1}

\subsection{Motivation}

Most of the models described above have the advantage of allowing 
for the inverse hierarchy of mass splittings which can enhance the
galactic 511 keV signal seen by INTEGRAL; however this comes at the
expense of a nonthermal history for the DM in order to keep the
intermediate mass state from being depopulated in the early universe.
Furthermore, purely SU(2) DM models do not allow for the excitation
$\chi_1\chi_1\to \chi_2\chi_3$ which would have half the energy
requirement of $\chi_1\chi_1\to \chi_3\chi_3$.  In the former case,
one need only produce a single $e^+ e^-$ pair (if $M_2$ is only
slightly above $M_1$), while in the latter, there must be at least
enough energy for two pairs, and the excitation rate is therefore
suppressed by the lack of sufficiently energetic DM particles in the
galactic center.  On the other hand, models with an extra U(1) in the
dark gauge sector can have  $\chi_1\chi_1\to \chi_2\chi_3$ by virtue
of mixing between the gauge groups when they are spontaneously broken.
%Furthermore, they allow the possibility of simultaneous explanation of
%the DAMA and INTEGRAL signals.

Even the simplest SU(2)$\times$U(1) model is considerably more
complicated than most of the pure SU(2) examples.  First, the DM is
necessarily vector-like (Dirac), in order to have a large bare mass
while carrying the extra U(1) charge, but the Dirac states must be
split into Majorana states by the Higgs which spontaneously breaks
the U(1).  Furthermore, the gauge group must be completely broken,
unlike the standard model where SU(2)$\times$U(1) breaks to $U(1)$.
Following \cite{Cheung} we refer to this extra requirement as
``charge breaking.''\ \   Custodial symmetry needs to also be broken
in order for the excited DM states to be able to decay into SM
particles, since otherwise the gauge bosons can be paired up into
charged states such as $W^\pm = \sqrt{1/2}(B_1\pm i B_2)$, analogous
to the $W$ bosons of the SM.  This charge is conserved if custodial
symmetry is unbroken [the ``charge breaking'' mentioned above only
insures that there is no unbroken U(1)], which would prevent the
transitions between similarly charged $\chi$ states needed by the XDM
and iDM  mechanisms.   Here we will analyze in some detail a model of
triplet SU(2)$\times$U(1) DM with these necessary properties,  which
was outlined in ref.\ \cite{Cheung}.  The potential needed for getting
the desired pattern of Higgs VEVs is presented there.  

\subsection{Specification of the model}

Consider two Weyl triplets $\chi_i$ and $\chi'_i$ which have
equal and opposite dark hypercharge $\pm y'/2$.  They can be given the
bare mass term $M\chi_i \chi'_i$.  Once the gauge symmetry is broken,
mass splittings can arise both through radiative corrections and
through Yukawa couplings to Higgs fields which acquire VEV's. We will
assume that the radiative corrections dominate the mass splittings
which respect $\chi-\chi'$ number conservation (the Dirac mass terms),
while the Yukawa couplings (see (\ref{yuk}) below) are responsible
for splitting the degenerate Dirac states into Majorana ones.  

To compute the radiative corrections, we must first find the spectrum
of gauge bosons.  As shown in ref.\ \cite{Cheung}, complete breaking
of SU(2)$\times$U(1) requires two Higgs doublets with equal and
opposite dark hypercharges $\pm y$, whose VEV's take the form
\beq
	h_1 = v_1 \left({\cos\alpha\atop\sin\alpha}\right),\quad
	h_2 = v_2 \left({0\atop 1}\right)
\eeq
For convenience we will also include a triplet Higgs field $\Delta_i$
with VEV $\langle \Delta_i\rangle = \Delta \delta_{i1}$. This breaks 
the custodial symmetry at tree level.  Without the triplet, custodial
symmetry breaking first appears at one loop.  To avoid the extra
effort of computing loop effects, we parametrize the symmetry
breaking using the triplet VEV.  Finally, it is necessary to include
a Higgs field $\phi$ with dark hypercharge $-y'$
 which can split the Dirac components of the DM states,
so that there are no diagonal DM couplings of the U(1) which mixes
with SM hypercharge; such couplings are strongly constrained by direct
DM searches, as discussed in section \ref{cdc}.   The Yukawa couplings
which accomplish this are
\beq
	h\phi\chi_i\chi_i + h'\phi^*\chi'_i\chi'_i
\label{yuk}
\eeq

\subsection{Mass eigenstates}

The VEV's of the four Higgs fields $h_i,\Delta,\phi$ give rise to
the gauge boson mass matrix in the basis $B_1,B_2,B_3,Y$
\beq
	 \left(\begin{array}{cccc} 
	 A  & 0 & 0 & g y s_{2\alpha} v_1^2\\
	0 &  A+ \delta & 0 & 0 \\
	0 & 0 &  A + \delta & gy(c_{2\alpha}v_1^2 +v_2^2)\\
	g y s_{2\alpha} v_1^2 & 0 & gy(c_{2\alpha}v_1^2 +v_2^2) &
	B
	\end{array}\right)
\label{gbmm}
\eeq
where 
$ A = g^2 v^2$, $B= y^2 v^2 + y'^2\phi^2$,
$v^2 = v_1^2+v_2^2$, $c_{2\alpha}=\cos 2\alpha$,
$s_{2\alpha}=\sin 2\alpha$,
$\delta = g^2\Delta^2$ and $\phi$ represents the VEV of the
U(1)-breaking Higgs field in (\ref{yuk}).   In order to give analytic
expressions, we will consider the off-diagonal charge-breaking 
elements to be small
perturbations $\epsilon_i$, and the custodial breaking to 
be even smaller, $\delta < \sum_i\epsilon_i^2 /| A-B|$.

The diagonalization of (\ref{gbmm}) is worked out in appendix
\ref{dmm}.  The masses eigenvalues of the four gauge bosons are given
in eqs.\ (\ref{mumasses}) and the mixings in eq.\ (\ref{rotmat}). 
The standard model couplings induced by the mixing term $\epsilon
Y_{\mu\nu}Y_{SM}^{\mu\nu}$  and the rotation matrix (\ref{rotmat})
thus involve the mass eigenstates $B_1'$, $B_3'$, $Y'$, while $B_2$
remains uncoupled to the SM currents.  As a result, all the gauge
bosons except $B_2$ have relative short lifetimes due to the decay
into $e^+ e^-$.  We come back to the decays of $B_2$ below.  

The aforementioned 
gauge interactions give rise to radiative corrections to the Dirac
masses of the form $\sum_i \chi_i \delta M_i \chi'_i$.  Only the
SU(2) gauge interaction vertices contribute to the splittings, 
because the hypercharge interactions give equal contributions to
each $\delta M_i$.  Relating the flavor eigenstates $B_i$ in 
terms of the mass eigenstates $B'_a$ by $B_i= R_{ia} B'_a$,
and denoting $B'_4 = Y'$, the contributions to the $\delta M_i$
are
\beqa
\delta M_1 &=& -\frac12\alpha_g \left(\mu_2 + \sum_i
R_{3i}^2\mu_i\right)\nonumber\\
\delta M_2 &=& -\frac12\alpha_g \sum_i
\left(R_{1i}^2+ R_{3i}^2\right)\mu_i \nonumber\\
\delta M_3 &=& -\frac12\alpha_g \left(\mu_2 + \sum_i
R_{1i}^2\mu_i\right)
\eeqa
The coefficients $R_{ia}$ are given in eq.\ (\ref{rotmat}).
Ignoring terms of $O(\delta^2)$, and subtracting the $\delta M_3$
contribution from all $\delta M_i$ (since we are only interested
in mass differences) we obtain
\beqa
\delta M_1 &=& \frac14\alpha_g \left({s_\theta^2\delta\over \sqrt{A}}
- \epsilon^2 f(A,B)\right) \nonumber\\
\delta M_2 &=& \frac14\alpha_g \left({\delta\over \sqrt{A}}
-2\psi^2\sqrt{A}  - \epsilon^2 f(A,B) \right) \nonumber\\
\delta M_3 &\equiv& 0
\label{split}
\eeqa
where
$\epsilon=\sqrt{\epsilon_1^2+\epsilon_2^2}$,
$\tan\theta = \epsilon_1/\epsilon_2$, $f(A,B) =$ $2\sqrt{B}
(A-B)^{-2}$ $+ [\sqrt{A}(A-B)]^{-1}$, and
$\psi = c_\theta s_\theta (A-B){\delta/\epsilon^2}$.  
One can show that the function $f(A,B)$ is positive for all values
of $A,B$.  
Since we have assumed that
$\delta \ll \epsilon^2/|A-B|$, this gives a normal hierarchy
with $M_1\sim M_2 < M_3$.  Whether $\chi_1$ or $\chi_2$ is the
lightest state depends on the $\psi^2$
term in $\delta M_2$.  It is of order $\delta^2/\epsilon^4$, 
which can
compete with the order $\delta$ term since only $\delta/\epsilon^2$
need be small for the consistency of our approximations.

The above mass splittings refer to Dirac states in the absence of
the Majorana masses induced by (\ref{yuk}) due to the VEV of $\phi$.
The effect of the latter is to split the Dirac mass eigenvalues
by $\pm \frac12(h+h')\langle\phi\rangle$.  Thus we get two sets of
states whose mass splittings are given by (\ref{split}), but they
are offset from each other by $m = (h+h')\langle\phi\rangle$.  It
would be consistent with our approximations to consider $m\gg \delta
M_i$, so that the more massive set of states does not play any
role at late times.

\subsection{Phenomenology}

It is interesting that the mass splittings (\ref{split}) are
parametrically suppressed to smaller values than the generic
$\alpha_g\mu$ estimate, and that the hierarchy between custodial
symmetry breaking and charge breaking, $\delta/\epsilon^2$, translates
into the hierarchy of masses $M_1\sim M_2 < M_3$, {\it i.e.,} it
explains why one mass splitting should be parametrically smaller than
the other.  In terms of the model parameters, the smallness of
$\epsilon_i$ can be arranged by assuming $y\ll g$, {\it i.e.,}
the dark hypercharge of the Higgs doublets is much smaller than the
SU(2) gauge coupling.  The even smaller breaking of custodial symmetry
seems to require an unnaturally small triplet VEV $\Delta$ in our
implementation, but as mentioned above, this is only a crutch to avoid
loop computations, since custodial symmetry is violated by hypercharge
interactions even without the triplet.  Because of the loop
suppression on top of the small $y$ coupling, this effect is indeed
expected to be smaller than the charge breaking.  

\begin{figure}[t] %\smallskip 
\centerline{\epsfxsize=0.4\textwidth\epsfbox{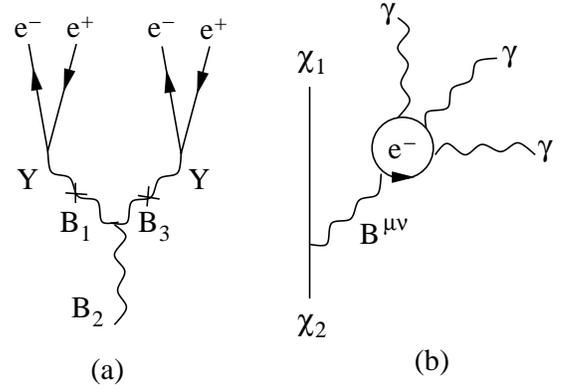}} 
\caption{(a) Left: diagram for $B_2\to 4e$ decay.
(b) Right: diagram for $\chi_2\to\chi_1+3\gamma$ decay.
} \label{loop3}
\end{figure}

Another striking point  is that $B_2$ is the only nonabelian gauge
boson which has no mixing with $Y$, due to the form of the mass
matrix (\ref{gbmm}), hence neither does  $B_2$ mix with the SM
hypercharge (however $B_2$ does decay into electrons via the process
shown in fig.\ \ref{loop3}(a)).  This means the transition between
$\chi_1$ and $\chi_3$ is the only one that does not couple to the
electron current, while the $\chi_2\lra \chi_1$ and $\chi_2\lra
\chi_3$ transitions do.  The $\chi_1\lra \chi_2$ transition can thus
be used for the iDM mechanism. Moreover, assuming $\chi_1$ is the
lightest state,  $\chi_1\chi_1\to\chi_3\chi_2$ (possible because of
$B_3$-$Y$ mixing) followed by $\chi_3\to\chi_2 e^+e^-$ can realize
the  XDM scenario, if $\delta M_{32} = M_3-M_2> 2m_e$.   The other
possibility is that $\chi_2$ is the lowest state, so that
$\chi_2\chi_2\to \chi_1\chi_3$ (enabled by $B_3$-$B_1$ mixing) is the
excitation channel.

In the case where $\chi_1$ is the ground state, we have the situation
where $\chi_2$ is a long-lived intermediate state whose
transitions $\chi_2\lra\chi_{1,3}$ couple to currents involving
nucleons or electrons.  This situation is highly constrained by
direct detection experiments, as we explained in section \ref{llis}.
Let us first show that $\chi_2$ is cosmologically long-lived.
The $\chi_2$ state can decay
to $\chi_1 + 3 \gamma$ through the operator 
$\epsilon \alpha^2 m_e^{-4}B_3 F^3$ which is induced through a 
virtual electron loop, shown in fig.\ \ref{loop3}(b). 
 Ref.\ \cite{BPR} (see also \cite{PRV}) estimated the rate to be 
\beq
	\Gamma(\chi_2\to \chi_1 +3\gamma) = {17 \epsilon^2\alpha_g\alpha^4 (\delta 
	M_{32})^{13} \over 2^7 3^6 5^3 \pi^3\, m_e^8\, \mu^4}
\eeq
For the parameter values we have favored, 
$\epsilon = 10^{-3}$, $M_\chi=1$ TeV, $\mu = 100$ MeV, 
$\delta M_{32} = 100$ keV, $\alpha_g = 0.04$, the lifetime is around
$10^{21}$ s, much greater than the age of the universe.  The
constraint (\ref{ddc}) thus applies.  According to ref.\ \cite{BPR},
such small couplings are inconsistent with an iDM explanantion of the
DAMA observations.  Thus the main potential advantage of this model
over a simpler one, such as doublet DM, is the lower energy threshold
excitation channel $\chi_1 \chi_1 \to \chi_2\chi_3$ for XDM.

\section{Conclusions}\label{conc}

We have surveyed a range of experimental and cosmological constraints
on the simplest models of dark matter with a hidden SU(2) gauge
symmetry, with a view toward explaining the PAMELA,  Fermi/LAT and
INTEGRAL/SPI electron/positron excesses by DM annihilation or
excitation.  Although new constraints on inverse Compton gamma rays
associated with the $e^+e^-$ production are making it more difficult
to accommodate the scenario,  models like those we pinpoint where
$M_\chi\cong 1$ TeV, $\alpha_g\sim 0.04$, the mass of the intermediate gauge or Higgs boson
is $\lsim 100$ MeV, and the annihilations proceed via 
$\chi\chi \to 4e$ rather than $2e$ or any combination of heavier
leptons, seem to still be viable.  

There are two uncertainties in the properties of DM halos which can
help alleviate the constraints from  gamma rays produced in our own
galaxy.  One is the possible presence of many subhalos with low
velocity dispersion being the principal regions of dark matter
annihilation would displace the gamma rays away from the galactic
center, where constraints from HESS are strongest.    The other
arises from  new studies of the effects of baryons on the DM velocity
dispersion profiles, which imply that Sommerfeld-enhanced
annihilation would be suppressed near the galactic center.  These two
effects could work together such that PAMELA/Fermi observations are
dominated by subhalo annihilations, while the the 511 keV excess is
enhanced by the larger DM velocities in the galactic center. We have
ruled out a third possibility in the present class of models, namely
that the intermediate particles decaying into $e^+e^-$ travel away
from the galactic center before decaying; we showed that such
particles would necessarily spoil BBN because of their long lifetimes
and high abundance in the early universe.  

All of these loopholes are relatively unimportant for a related class
of constraints, which considers the gamma rays emitted by all halos
at all redshifts, including their effect on the CMB.  It would be
important to reconsider these constraints specifically for
annihilations in which $\chi\chi \to 4e$, and higher numbers of
electrons/positrons, due to cascading of the dark gauge bosons, to
see how much they really constrain the present class of models.  

We have shown that there is a new potential signal which could
provide additional evidence for nonabelian DM, if the XDM
interpretation of the INTEGRAL 511 keV excess is correct.  The DM
transition magnetic moment interaction induced at one loop, due to
the nonabelian terms in the gauge kinetic mixing, should give rise to
a narrow gamma ray line with energy equal to the mass splitting of
the two DM states, expected to be of order a few MeV. Our estimates
show that for $\alpha_g\sim 0.04$, the strength of this line can
be close to the current sensitivity of INTEGRAL/SPI, if the phase
space for excited dark matter decay is accidentally small.

Another potential signal is through the couplings of the portal
particle which interacts with the SM, with coupling reduced by the
factor $\epsilon$ if it is a dark gauge boson, or mixing angle
$\theta$ if it is a Higgs boson.  The $\epsilon$ factor is already
strongly constrained, depending on the mass $\mu$ of the portal
boson, by precision QED tests and beam dump experiments.  Proposed
fixed target experiments could further probe the allowed range of
$\epsilon$ in the near future. We have derived the bound $\epsilon
\gsim
4\times 10^{-11}$ in the case of gauge kinetic mixing,  from the
requirement that dark gauge bosons annihilate to $e^+e^-$ efficiently
before nucleosynthesis.  If any of the gauge bosons are stable, we
get the stronger bound $\epsilon \gsim 3\times 10^{-10}$ from
overclosure.  This is three orders of magnitude
smaller than current limits from the E137 beam dump experiment. In the case of Higgs mixing,
we find an analogous lower bound $\theta\gsim 10^{-6}$ on the Higgs
mixing angle.

We have also shown that Higgs mixing rather than gauge kinetic mixing
can be a viable portal to the SM, despite first appearances that it
would tend to induce too high a rate of $\chi_i\chi_i\to f\bar f$,
where $f$ is any SM fermion, preferably the top quark.  Such diagonal
couplings of the Higgs to $\chi_i\chi_i$ might also be expected to 
lead to too high
of a cross section for scattering from nucleons, $\chi N\to \chi N'$,
in direct DM detectors.  We have shown that in fact the constraints
can easily be satisfied for reasonable values of the couplings.

Concerning specific models, the simplest possibility is DM in the
doublet representation, which has the potential for implementing the
XDM mechanism to explain INTEGRAL/SPI observations of 511 keV gamma
rays from the galactic center.   We noted that the doublet model must
get its mass splitting from a Yukawa coupling to a triplet Higgs,
rather than from radiative corrections, which divorces the scale of
the mass splitting from the strength of the gauge coupling.

A recurring theme of our paper was the possibility of an inverted
mass hierarchy in models with three or more DM components, which can
help boost the production of the 511 keV excess by the XDM
mechanism.  A $Z_2$ discrete symmetry is required to keep the
intermediate-mass  DM state stable in this scenario.  We showed that
models of triplet and quintuplet DM afford many examples with the
desired properties, as well as the alternative normal mass
hierarchy.   In both triplet and quintuplet DM models, it is possible
to choose Higgs VEVs such that the mass splittings are smaller than
the generic estimate $\alpha_g\mu$ for the size of the radiative
corrections, which can be help explain the hierarchy between
splittings needed for the iDM or inverse mass hierarchy XDM
scenarios. An in-depth reanalysis of the viability of the XDM
mechanism for explaining the 511 keV observation, in the context of
the present class of models, is in progress \cite{new}.

We have not focused on the DAMA/LIBRA annual modulation as one of the
signals to be explained by nonabelian dark matter, even though that
was cited as one of the original motivations for this class of
models.  Our choice stems from recent results
\cite{dama-constraints2} which note that the iDM explanation for the
signal is ruled out at the 99\% c.l.\ by data from the ZEPLIN-II and 
at 95\% c.l.\ by XENON10 and CRESST II observations, for DM whose
mass is in the range of interest for explaining the PAMELA/Fermi
lepton excesses.  Another reason is that our proposal of the inverted
mass hierarchy for boosting the XDM mechanism is at odds with the
normal mass hierarchy needed for iDM.  However if one believes there
is still room for iDM to work, then  the desired mass splittings can
be achieved within the models considered here (for DM in triplet and
higher representations), since there is great freedom to adjust
the DM spectra through ratios of the VEV's of the dark Higgs fields.

In summary, SU(2) gauge theories of DM continue to offer an elegant
explanation for numerous effects, with intricate implications for
cosmology, DM halo properties, laboratory tests, and the prospect
for being ruled in or out in the near future.

\bigskip

{\bf Acknoweldgment.}  We thank Sabine Kraml for assistance with
MicrOMEGAs, Guy Moore for information about the Higgs-nucleon
coupling, Emiliano Mocchiutti for correspondence about the PAMELA
antiproton background, Julio Navarro for discussions on the effect of baryons on
DM halos, Maxim Pospelov for discussions about long-lived intermediate
states,  Andrew Strong for valuable advice about the
interpretation of INTEGRAL/SPI data, and Aaron Vincent for discussions
about GALPROP.  Our work is supported by NSERC of Canada.

\appendix
\section{Transition magnetic moment}
\label{tmm}
By routing the external momenta through the loop in the appropriate
way, the expression for the loop diagram which gives rise to the
transition magnetic moment can be written as
\beq
	\int{d^4p\over (2\pi)^4} {\epsilon g^3 \,\gamma_\mu (\slashed{p} + 
	\slashed{\bar q} + M_1)\gamma_\nu\over
	[(p+\bar q)^2 - M_1^2][(p+\delta q)^2-\mu^2]
	[(p-\delta q)^2-\mu^2]}
\eeq
antisymmetrized over $\mu,\nu$ (since it is contracted with
$F_{\mu\nu}$ of the external photon), where $\bar q$ is the average
4-momentum of the two external DM states, $\bar q = \frac12(q_3+q_2)$,
and $\delta q$ is half the 4-momentum of the photon, $\delta q = 
\frac12(q_3-q_2)$, and $M_1$ is the mass of the virtual DM particle.
We have ignored mass differences between the two
gauge bosons in the loop since this has a subleading effect on the
result.  Using Feynman parameters and Wick rotating, the $p$ integral
can be done, leading to 
\beqa
&&\epsilon g^3\, \int_0^1 \!\!\!dx\int_0^{1-x}\!\!\!\!\!\!\!\!\! dy  \\
	&&{\gamma_\mu((x+y)\slashed{\bar q} + (y-x)\delta \slashed{q}
	+ M_1)\gamma_\nu\over
	z^2 M^2 + \frac12(y-x)z \delta M_{32}^2
	+\delta M^2_{123}z + \mu^2(x+y)} \nonumber
\eeqa
where $z=1-x-y$, $\delta M^2_{123}= M_1^2-\frac12(M_2^2+M_3^2)$,
and $\delta M_{32}^2=M_3^2-M_2^2$.  For the parameter values of interest,
we find that it is a good approximation to set 
$\delta M_{32}^2=\delta M^2_{123}=0$ in the denominator.  By
anticommuting gamma matrices in the numerator and using the Dirac
equation for the external spinors, one can show that $\slashed{\bar q}
\to -\frac12(M_2+M_3)$, while $\delta \slashed{q}$ gives a subleading
in $\delta M_{23}$ contribution which can be neglected.  Furthermore,
it is a good approximation to set $(x+y)=1$ for the coefficient of
$\mu^2$ in the denominator.  In this way one can get the analytic
approximation (\ref{MM}), which we have numerically verified to be
good in the range of parameters of interest. 

\section{Radiative DM mass splittings}
\label{rmca}
In this appendix we present the radiative mass corrections to a DM
multiplet $\chi_i$ by virtual massive gauge bosons, as shown in 
fig.\ \ref{self-energy}(a).
%\begin{figure}[t]
%\centerline{\input{massplit.pstex_t}} \caption{Radiative correction
%to mass splitting within the DM multiplet}\label{Amassplit}
%\end{figure}
Assume the DM multiplet transforms under a gauge group $G$ with
generators $T_j$. Also assume the gauge bosons $A_j$ have mass
$\mu_j$. Figure \ref{self-energy}(a) gives a correction to the 
self-energy of $\chi_i$ of
\begin{eqnarray}
\delta M_i=g^2\sum_{j,a}T_{ia}^jT_{ai}^j\int_0^1 dx \int
\frac{d^4k_E}{(2\pi)^4}\;\frac{4M_a-2\slashed{p}x}{(k_E^2+\Delta)^2}\label{A1}
\end{eqnarray}
where $\Delta=-M_i^2(1-x)x+\mu_j^2x+M_a^2(1-x)\simeq
M_a^2(1-x)^2+\mu_j^2x$. After integrating over the Euclidean
4-momentum $k_E$ and using the equation of motion to set
$\slashed{p}\to M_i$, 
 we find two pieces, one of which is ultraviolet (UV)
 divergent, 
while the other is infrared (IR) divergent as $\mu_j\to 0$,
\begin{eqnarray}
\delta M_i=\frac{g^2}{8\pi^2}\sum_{j,a}T_{ia}^jT_{ai}^j\int_0^1 dx
\;(2M_a-M_i x)\;(\ln\Lambda^2-\ln\Delta)\nonumber\\
\label{A2}
\end{eqnarray}
where $\Lambda$ is the ultraviolet cutoff of $k_E$. We are only
interested in the IR divergent term  because the UV divergent term
cancels out when considering mass splittings. Notice that the IR 
divergence
occurs when $x\to 1$, so we can set $x=1$ in the first factor
of the integrand.

Further assuming that the gauge boson mass is much smaller than 
the DM mass, $\mu_j\ll M_a$, and ignoring the $\chi$ mass splittings
on the r.h.s., the IR divergent term turns out to be
\begin{eqnarray}
\delta M_i&=&-\frac{\alpha }{2\pi}M_\chi\sum_{j,a}
T_{ia}^jT_{ai}^j\int_0^1 dx\nonumber\\
&&\;\left(\ln\left(1+\frac{x\mu_j^2}{(1-x)^2M_\chi^2}\right)
+2\ln\left((1-x)M_\chi\right)\right)\nonumber\\
&\rightarrow&-\frac{\alpha}{2}\sum_{j,a}\mu_jT_{ia}^jT_{ai}^j\label{A3}
\end{eqnarray}
We dropped the last term because it is the same for each component
of the DM multiplet and thus has no effect on the mass splitting.

Now we turn to the mass correction from virtual Higgs bosons.
For illustration, suppose we have triplet DM $\chi_i$ coupled to a
quintuplet scalar $S_{ij}$, via $h_s\chi_i S_{ij}\chi_j$. This
induces radiative mass corrections through the diagram of
fig.\ \ref{self-energy}(b).  Similarly to the gauge boson case we
have,
\begin{eqnarray}
\delta M_i&=&\frac{\alpha_{h} }{4\pi}\sum_j M_j
\int_0^1 dx\nonumber\\
&&\;\left(\ln\left(1+\frac{x\, m_{ij}^2}{(1-x)^2M_j^2}\right)
+2\ln\left((1-x)M_j\right)\right)\nonumber\\
&\rightarrow&\frac{\alpha_{h}}{4}\sum_j m_{ij}\label{A4}
\end{eqnarray}
where the last approximation applies 
if the dark Higgs boson mass is much smaller than the dark matter
mass, $m_{ij}\ll M_a$.

\section{DM annihilation cross section}
\label{AppC}
Here we derive the annihilation cross sections for DM in the
doublet, triplet and quintuplet representations.
Because of its relative simplicity we start with the triplet 
case, assuming Majorana DM, whose gauge interaction Lagrangian is
\beq
	{\cal L}_{\rm int} = \frac12\, g\,\epsilon_{abc}\,\bar\chi_a B_\mu^b\gamma^\mu \chi_c
\eeq 
For the annihilation channel $\chi_1\chi_1\to B_2 B_2$, the relevant
interaction is
\beq
	g \bar \chi_1 B_2^\mu\gamma_\mu \chi_3
\eeq
where the antisymmetric property of the  Majorana vector current,
$\bar\chi_3\gamma^\mu\chi_1 = - \bar\chi_1\gamma^\mu\chi_3$ has
been used.  The matrix element for $\chi_1\chi_1\to B_2 B_2$ is then
\beqa
\!\!\!\!\!\!\!\!\!\!\!\!\!{\cal M}_{11} &=& 	-i g^2\bar v(p_1) \slashed{\epsilon}(q_1) {1\over
	\slashed{p}_1 - \slashed{q}_1 - M_\chi} \slashed{\epsilon}(q_2)
	u(p_2) \nonumber\\
	&&-i g^2\bar v(p_1) \slashed{\epsilon}(q_2) {1\over
	\slashed{p}_1 - \slashed{q}_2 - M_\chi} \slashed{\epsilon}(q_1)
	u(p_2)
\eeqa 
where $p_i$ are the incoming $\chi$ momenta and $q_i$ are the outgoing
$B$ momenta, and the $\epsilon$'s are polarization vectors.  This has
algebraically the same form as the matrix element for
electron-positron annihilation.  In the nonrelativistic and 
$M_\chi\gg\mu$
limits, the spin-averaged squared matrix element is $2g^4$ for
both the $t$ and $u$ channels squared.  The interference term vanishes
in these limits.  The sum over final states includes $\chi_1\chi_1\to
B_3 B_3$ as well, so the total is $8 g^4$.

Next we consider the $\chi_1\chi_2\to B_1 B_2$, 
which procedes by a sum of $t$-channel mediated by internal $\chi_3$
and $s$-channel mediated by $B_3$ (using the 3-gauge boson vertex).
The matrix element is
\beqa	
\!\!\!\!\!\!\!\!\!\!\!\!\!{\cal M}_{12} &=&
-i g^2\bar v(p_1) \slashed{\epsilon}(q_1) {1\over
	\slashed{p}_1 - \slashed{q}_1 - M_\chi} \slashed{\epsilon}(q_2)
	u(p_2) \nonumber\\
	&-&i g^2{\bar v(p_1) \gamma_\nu u(p_2)\over
	p_s^2 - \mu^2}\left[ \eta_{\nu\lambda}(p_s+q_2)_\mu
	\right.\nonumber\\
	&&\qquad \left.-\eta_{\nu\mu}(p_s+q_1)_\lambda + 
	\eta_{\mu\lambda}(q_1-q_2)_\nu\right]
\eeqa 
where $p_s=p_1+p_2$.  	We find that the spin-averaged squared
matrix element gets contributions of $2 g^4$, $-(19/4) g^4$
and $4 g^4$ from the $t^2$, $s^2$ and interference channels,
respectively, in the same limits as mentioned above.  
(The fact that the direct $s^2$ term is negative
is due to the unphysical polarizations in the sum over final state
gauge bosons; only the full amplitude squared is physically
meaningful.)  The total is thus $(5/4) g^4$.  

Finally we must average over the initial colors to give
$\langle|{\cal M}^2|\rangle = \frac13 {\cal M}_{11}^2 +
\frac23 {\cal M}_{12}^2 = [8/3 + (2/3)(5/4)]g^4 = (7/2)g^4$.
This must be multiplied by an additional factor of $1/2$ for the indistinguishability
of the final states.
The differential cross section is thus 
\beq
	{d\sigma\over dt} = {(7/4) g^4\over 64\pi s M_\chi^2 v^2}
\eeq
evaluated in the center of mass frame.  Integrating over the range of the
Mandelstam $t$ variable $\delta t = 4 M^2 v$ and using the 
relative velocity $v_{\rm rel} = 2v$, we obtain the cross section
(\ref{sigann}) for the triplet case.

To find the annihilation cross section for DM in other
representations, we work out the group theory factors for the general
case.  For the $t^2$ and $u^2$ terms, these take the form
$(1/d_R^2)\sum_{ab}$tr($T^a T^a T^b T^b) = C_2(R)^2/d_R$, where $d_R
= 2j+1$ is the dimension of the spin $j$ representation and $C_2(R) =
j(j+1)$ is the quadratic Casimir invariant.  The factors $(1/d_R^2)$
come from averaging over the colors of the initial states. For the
$s^2$ term we get  $(1/d_R^2)\sum_{ij}\sum_{abcd}
\epsilon^{abc}\epsilon^{abd} T^c_{ij} (T^d_{ij})^* =
2/d^2_R\sum_c{\rm tr}[T^c T^c] = 2 C_2(R)/d_R$.  For the $st$ term, we
find $(1/d_R^2)$tr$(T^a T^b T^c) \epsilon_{abc} = 
(i/2d_R^2)\epsilon_{abd} \epsilon_{abc} {\rm tr}(T^d T^c)\sim
C_2(R)/d_R.$

Using these results, we can generalize the triplet computation
to other representations.  Using $t^2$, $u^2$, $s^2$ and $st$ to
represent the contribution to the squared matrix element from any
definite external states (hence encoding the spin sums but not the
group theory factors) we have
\beq
	\langle|{\cal M}^2|\rangle = \frac23 \left(\frac32(t^2+u^2)
	+ s^2 + st \right)
\eeq
in the triplet representation, which based on the above group theory
factors, generalizes to
\beq
	\frac23 
	{3\over d_R}\left(\frac32\left({C_2(R)\over 2}\right)^2(t^2+u^2)
	+ {C_2(R)\over 2}(s^2 + st) \right)
\eeq
where $t^2=u^2 = 2 g^4$, $s^2 = -(19/4) g^4$ and $st = 4 g^4$.
We obtain
\beq
	\langle|{\cal M}^2|\rangle = {3 j (j+1)\over 2 j + 1}
	\left[ j(j+1) - \frac14 \right]g^4
\eeq
for the spin-$j$ representation.  This must still be multiplied
by the symmetry factor $1/2$ for identical final states.

\section{Rate of $3B\to 2B$ down-scattering process}
\label{3b2b}

In the case of $\epsilon\ll 1$ where dark gauge bosons do not stay in
kinetic equilibrium with the SM particles, the processes which can
delay  them from dominating the energy density of the universe after
becoming nonrelativistic are those which convert 3 to 2 particles.
The squared matrix element for $3B\to 2B$ in nonabelian gauge theory
is of order $|{\cal M}|^2 \sim g^6/\mu^2$ if the $B$'s are
nonrelativistic.  The Boltzmann equation for the number density of 
$B$'s takes the form $\dot n_B + 3Hn_B = {\cal C}$.  The collision
term of interest for  $3B\to 2B$ is
\beqa
	{\cal C}&\sim& -\int \prod_i {3\, d^3 p_i\over (2\pi)^3 2 E_i}
	f_1 f_2 f_3
 |{\cal M}|^2 \nonumber\\
	&\times& (2\pi)^4 \delta^{(4)}(p_1 + p_2 + p_3 - p_4 - p_5)
	\nonumber\\
	&\cong& {n_B^3(T)(3/2)^5(4\pi)^4 \over (2\pi)^{11} \mu^3}
	{g^6\over\mu^2}
\eeqa
where the nonrelativistic density is $n_B(T) = (\mu T)^{3/2}
e^{-\mu/T}$.  To convert this to a rate, we should divide by one power
of $n_B$.  Equating this to the Hubble rate, we find that
\beq
	2x_f = \ln\left({3.4\alpha_g^3 M_p\over \mu}\right) -\ln x_f
\eeq
leading to the result in the text above eq.\ (\ref{yb1eq}).

\section{Diagonalization of SU(2)$\times$U(1) model mass matrices}
\label{dmm}
We give the details of approximately solving for the mass eigenvalues
and eigenstates in the SU(2)$\times U(1)$ DM model discussed in 
section \ref{su2u1}.  The dark gauge bosons are denoted by $B_i$
and $Y$, and the effects of mixing with the SM hypercharge are
neglected in the following approximations.  
 Since $B_2$ does not mix with the other fields, 
we can consider the nontrivial 3$\times$3 mass matrix in 
the $B_1,B_3,Y$ basis, writing it in the form
\beq
	\mu^2 = \left(\begin{array}{cc}\bar  A & \epsilon\\ \epsilon^t & B
	\end{array}\right)
\eeq
where $\bar A= $ diag$(g^2 v^2,g^2 v^2 + \delta)$
\beq
	\epsilon = 
gy\left({s_{2\alpha} v_1^2\atop c_{2\alpha}v_1^2 +v_2^2}
\right) = \left({\epsilon_1\atop\epsilon_2}\right)
\eeq
and $B = y^2 v^2 + y'^2\phi^2$.  
It is convenient to change bases using a global SU(2) rotation
which makes only the lower component of $\epsilon$ nonzero.
This is accomplished using a $2\times 2$ rotation $R(\theta)$
in the $B_1$-$B_3$ subspace, with $\tan\theta = \epsilon_1/\epsilon_2$.  In the
new basis, $\bar A$ is no longer diagonal,
\beq
	\bar A \to g^2 v^2 {\bf 1} + \delta \left(\begin{array}{cc}
 	 s^2_\theta	& c_\theta s_\theta \\
	c_\theta s_\theta & 	 c^2_\theta 
	\end{array}\right)
\eeq
where $s_\theta =\sin\theta$, $c_\theta = \cos\theta$,
and ${\bf 1}$ is the 2$\times$2 unit matrix
Treating $\epsilon$ as a perturbation, one can perform a 3$\times$3
rotation to get rid of the off-diagonal $\epsilon$ blocks, using
\beq
	O = \left(\begin{array}{cc}
	{\bf 1} - \eta \eta^t/2	&	-\eta \\
	\eta^t			& 1 - \eta^t \eta/2
	\end{array}\right)
\eeq
where $\eta = (\bar A - B\times{\bf 1})^{-1}\epsilon 
\equiv \tilde A^{-1}\epsilon$.  Under this rotation,
the 2$\times$2 block $\bar A$ transforms again, receiving a correction
$\bar A\to \bar A + \delta \bar A$ of the form
\beq
\delta 
\bar A = -\frac12\{\bar A ,\tilde A^{-1}  X \tilde A^{-1}\} 
	+ \{X,\tilde A^{-1} \} + B \tilde A^{-1}  X \tilde A^{-1} 
\eeq
where  $X$ is the 2$\times$2 matrix $\epsilon\epsilon^t$,
whose only nonvanishing component is $\epsilon^2$ in the 2,2 position.
We find that to leading order in $\delta$, 
\beq
	\delta \bar A = \left(\begin{array}{c|c}
	0 & -{c_\theta s_\theta\delta\epsilon^2\over 2(A-B)^2}\\
	\hline
	-{c_\theta s_\theta\delta\epsilon^2\over 2(A-B)^2}&
	{\epsilon^2\over A-B}\left(1-{c_\theta^2\delta\over 2
\label{dA}
	(A-B)}\right)\end{array}\right)
\eeq
where $A = g^2 v^2$.
$B$ gets a similar correction, but the correction to $\bar A$ is more
important since this splits the gauge boson mass eigenvalues, whereas
$B$ is already well separated from the eigenvalues of the $\bar A$ matrix.
The final step for the gauge boson mass eigenvalues is to diagonalize
$\bar A + \delta \bar A$.  The off-diagonal elements of 
$\delta \bar A$ give only $O(\delta^2)$ corrections to the gauge
boson  mass
splittings, which we are ignoring; thus, 
 denoting $A = g^2 v^2$ and 
$B = y^2 v^2 + y'^2\phi^2$, 
the resulting masses are
\beqa
	\mu_1^2 &=& A +  s_\theta^2 \delta + O(\delta^2)\\
 	\mu_2^2 &=& A + \delta \\
 	\mu_3^2 &=& A  + {\epsilon^2\over
	A - B}  + O(\delta)\\
 	\mu_4^2 &\cong& B
\label{mumasses}
\eeqa
The relation between the flavor (unprimed) and mass (primed)
eigenstates is $B_2 = B_2'$ and 
\beq
\left(\begin{array}{c}B_1\\B_3\\Y\end{array}\right)
\cong\left(\begin{array}{ccc} 1 & \psi & 0\\
	-\psi & 1 & -\eta\\
	-\eta\psi & \eta & 1\end{array}\right)
\left(\begin{array}{c}B_1'\\B_3'\\Y'\end{array}\right)
\label{rotmat}
\eeq
where
\beq
	\eta = {\epsilon\over A+\delta -B}, \quad
	\psi = c_\theta s_\theta (A-B){\delta\over \epsilon^2}
\eeq
Here $\psi$ is the small angle of the rotation which diagonalizes
 (\ref{dA}).  It gives rise to $O(\delta^2/\epsilon^4)$ contributions
to the mass splittings of the DM states, which are larger than the
$O(\delta^2)$ terms we have ignored thus far.
These formulas assume $\delta <\epsilon^2/|A-B|$, so they are not
valid in the limit $\epsilon\to 0$.

\end{document}